\documentclass{LMCS}

\usepackage{xspace}
\usepackage{alltt}
\usepackage{hyperref}
\usepackage{longtable}
\usepackage{enumerate}

\newcommand*{\ie}{i.e.,\xspace}

\newcommand*{\cf}{cf.\xspace}
\newcommand*{\eg}{e.g.\xspace}
\newcommand*{\resp}{resp.\xspace}
\newcommand*{\role}{r\^{o}le\xspace}
\newcommand*{\CIC}{\textrm{CIC}\xspace}
\newcommand*{\coq}{\textit{Coq}\xspace}
\newcommand*{\haskell}{\textit{Haskell}\xspace}
\newcommand*{\martinlof}{Martin-L\"of\xspace}

\newcommand*{\pavlovic}{Pavlovi\'{c}\xspace}

\newcommand*{\gimenez}{Gim\'{e}nez\xspace}
\newcommand*{\mobius}{M\"obius\xspace}
\newcommand*{\Setcoq}{\ensuremath{\textsf{Set}}\xspace}
\newcommand*{\Propcoq}{\ensuremath{\textsf{Prop}}\xspace}
\newcommand*{\Typecoq}{\ensuremath{\textsf{Type}}\xspace}
\newcommand*{\Sets}{\ensuremath{\textbf{Set}}\xspace}  
\newcommand*{\after}{\mathop{\circ}}            
\newcommand*{\afterM}{\ensuremath{\mathop{\circ}}\xspace}                    
\newcommand*{\afterL}{\ensuremath{\mathop{\bullet_1}}\xspace}                
 \newcommand*{\afterR}{\ensuremath{\mathop{\bullet_2}}\xspace}                
\newcommand*{\map}[3]{\ensuremath{#1\colon #2\longrightarrow #3}\xspace}
\newcommand*{\appl}{\ensuremath{\quad\!}}                               
\newcommand*{\arr}{\ensuremath{\mathord{\to}}\xspace}                   
\newcommand*{\set}[1]{\ensuremath{\{\,#1\,\}}\xspace}
\newcommand*{\pair}[2]{\ensuremath{\langle#1,#2\rangle}\xspace}
\newcommand*{\filter}{\ensuremath{\mathsf{filter}}\xspace}            
\newcommand*{\homographic}{\ensuremath{\mathsf{homographic}}\xspace}  
\newcommand*{\quadratic}{\ensuremath{\mathsf{quadratic}}\xspace}      
\newcommand*{\cofix}{\ensuremath{\textsf{cofix }}\xspace}             
\newcommand*{\hd}{\ensuremath{\mathsf{hd}}\xspace}            
\newcommand*{\tl}{\ensuremath{\mathsf{tl}}\xspace}            
\newcommand*{\sumbool}{\ensuremath{\mathop{\oplus}}\xspace}            
\newcommand*{\bisim}{\ensuremath{\mathop{\cong}}\xspace}
\newcommand*{\MAT}{\ensuremath{\mathbb{M}}\xspace}
\newcommand*{\TEN}{\ensuremath{\mathbb{T}}\xspace}
\newcommand*{\Ibase}{\ensuremath{[-1,1]}\xspace}
\newcommand*{\LL}{\ensuremath{\mathbf{L}}\xspace}
\newcommand*{\RR}{\ensuremath{\mathbf{R}}\xspace}
\newcommand*{\MM}{\ensuremath{\mathbf{M}}\xspace}
\newcommand*{\DIG}{\ensuremath{\mathbf{DIG}}\xspace}
\newcommand*{\DIGinf}{\ensuremath{\mathbf{DIG}^{\omega}}\xspace}
\newcommand*{\refining}[1]{\ensuremath {\operatorname{\mathbf{Ref}(#1)}}\xspace}
\newcommand*{\bounded}[1]{\ensuremath {\operatorname{\mathbf{Bounded}(#1)}}\xspace}
\newcommand*{\refiningvoid}{\ensuremath {\operatorname{\mathbf{Ref}}}\xspace}
\newcommand*{\boundedvoid}{\ensuremath {\operatorname{\mathbf{Bounded}}}\xspace}
\newcommand*{\emitting}[2]{\ensuremath {\operatorname{\mathbf{Incl}(#1,#2)}}\xspace}
\newcommand*{\emittingdec}[2]{\ensuremath{\operatorname{\mathbf{Incl}_{dec}(#1,#2)}}\xspace} 
\newcommand*{\emittingvoid}{\ensuremath {\operatorname{\mathbf{Incl}}}\xspace} 
\newcommand*{\emittingdecvoid}{\ensuremath{\operatorname{\mathbf{Incl}_{dec}}}\xspace} 
\newcommand*{\diameter}[1]{\ensuremath {\operatorname{\mathbf{diam}(#1)}}\xspace}
\newcommand*{\tdiameter}[1]{\ensuremath {\operatorname{\mathbf{diam}_2(#1)}}\xspace}
\newcommand*{\NN}{\ensuremath{\mathbb N}\xspace}
\newcommand*{\ZZ}{\ensuremath{\mathbb Z}\xspace}
\newcommand*{\QQ}{\ensuremath{\mathbb Q}\xspace}
\newcommand*{\RRR}{\ensuremath{\mathbb R}\xspace}
\newcommand*{\Streams}{\textsf{Streams}\xspace}
\newcommand*{\textcons}{\textsf{\upshape Cons}\xspace}
\newcommand*{\rep}{\textsf{\upshape rep}\xspace}
\newcommand*{\cons}{\ensuremath{\colon\!\!\!\colon}\xspace}
\newcommand*{\scottbracketl}{\ensuremath{\mathopen{[\![}}}
\newcommand*{\scottbracketr}{\ensuremath{\mathclose{]\!]}}}
\newcommand*{\decode}[1]{\ensuremath{{\scottbracketl{#1}\scottbracketr}}\xspace}
\newcommand*{\redun}[1]{\ensuremath{\operatorname{\mathsf{red}}(#1)}\xspace}
\newcommand*{\redunvoid}{\ensuremath{\operatorname{\mathsf{red}}}\xspace}
\newcommand*{\ppif}{\textbf{\upshape{ if }}\xspace}
\newcommand*{\ppelseif}{\textbf{\upshape{ else if }}\xspace}
\newcommand*{\ppow}{\textbf{\upshape{ otherwise.}}\xspace}
\newcommand{\judgement}[2]{\frac{\displaystyle\strut{#1}}{\displaystyle\strut{#2}}}
\newcommand{\slft}[4]{\left[\begin{smallmatrix}
                              #1&#2\\
                              #3&#4\\
                            \end{smallmatrix}\right]}
\newcommand{\stensor}[8]{\left[ \begin{smallmatrix}
                                #1&#2&#3&#4\\
                                #5&#6&#7&#8\\
                                \end{smallmatrix} \right]}

\newenvironment{coqtt}  
 {\fontfamily{cmtt}\selectfont\noindent\hrulefill\vspace{-2\smallskipamount}\begin{quote}\begin{alltt}}
 {\end{alltt}\end{quote}\vspace{-4\smallskipamount}\noindent\hrulefill\par\medskip\ignorespacesafterend%
}

\newcounter{tableau}

\def\doi{4 (3:6) 2008}
\lmcsheading%
{\doi}
{1--40}
{}
{}
{Jun.~11, 2007}
{Sep.~10, 2008}
{}   

\begin{document}

\title[Coinductive Formal Reasoning in Exact Real Arithmetic]{Coinductive
  Formal Reasoning in Exact Real Arithmetic\rsuper *}

\author[M.~Niqui]{Milad Niqui\rsuper a}
\address{Department of Software Engineering\\ Centrum Wiskunde \& Informatica\\ Kruislaan~413\\ NL-1098~SJ\\ Amsterdam\\ The Netherlands. {Tel: +31 20 5924073\\ Fax: +31 20 5924200.}}
\email{M.Niqui@cwi.nl}
\thanks{{\lsuper a}Former address: Institute for Computing and Information Sciences, Radboud
  University Nijmegen, The Netherlands. Research supported by the Netherlands
  Organisation for Scientific Research (NWO)}

\titlecomment{{\lsuper *}Parts of this work have appeared in \cite{milad.06a,milad.07c}.}
%%%%%%%%%%%%%%%%%%%%%%%%%%%%%%%%%%%%%%%%%%%%%%%%%%%%%%%%%%%%%%%%%%%%%%%%%%%%%%%%

\begin{abstract}
  \noindent    
  In this article we present a method for formally proving the correctness of
  the lazy algorithms for computing homographic and quadratic transformations
  --- of which field operations are special cases--- on a representation of real
  numbers by coinductive streams. The algorithms work on coinductive stream of
  \mobius maps and form the basis of the Edalat--Potts exact real arithmetic.
  We use the machinery of the \coq proof assistant for the coinductive types to
  present the formalisation. The formalised algorithms are only \emph{partially
    productive}, \ie they do not output provably infinite streams for all
  possible inputs. We show how to deal with this partiality in the presence of
  syntactic restrictions posed by the constructive type theory of \coq.
  Furthermore we show that the type theoretic techniques that we develop are
  compatible with the semantics of the algorithms as continuous maps on real
  numbers. The resulting \coq formalisation is available for public download.
\end{abstract}

\maketitle

%%%%%%%%%%%%%%%%%%%%%%%%%%%%%%%%%%%%%%%%%%%%%%%%%%%%%%%%%%%%%%%%%%%%%%%%%%%%%%%%
%                                                                              %
%                                Introduction                                  %
%                                                                              %
%%%%%%%%%%%%%%%%%%%%%%%%%%%%%%%%%%%%%%%%%%%%%%%%%%%%%%%%%%%%%%%%%%%%%%%%%%%%%%%%
\section*{Introduction}
Exact real numbers constitute one of the prime examples of infinite objects in
computer science.  The ubiquity and theoretical importance of real numbers as
well as recent safety-critical applications of exact arithmetic makes them an
important candidate for applying various approaches to formal verification.
Among such approaches one that is tailor-made for infinite objects is
\emph{coinductive reasoning}. A careful coinductive formalisation of real
numbers has two advantages: (1) it provides a certified package of exact
arithmetic; (2) it gives valuable insight into various notions of coinductive
proof principles that can contribute to the area of formal verification for
infinite objects.

Coinductive reasoning is dual to the usual approach of using algebraic and
inductive data types both for computation and reasoning and can be studied from
a set theoretical~\cite{barwise.96}, category theoretical~\cite{jacobs.97}, or
type theoretical~\cite{coquand.94} point of view. In all these settings the
coinductive structure of real numbers is usually expressible as \emph{streams}
which have a simple and well-understood shape. Although there are other
coinductive objects (\eg \emph{expression trees}~\cite{edalat.97}) modelling
exact real numbers, the stream approach has proven to be expressible enough for
most computational purposes.  In this approach a real number $r$ is represented
by a stream of nested intervals whose intersection is the singleton $\{r\}$.
This approach has always been the basis of representing real numbers, as the
usual decimal representation is an instance of this representation with digits
denoting interval-contracting maps.  Because of this, much work has been done in
the study and implementation of various algorithms for specific stream based
representations.  In this context one rather generic approach is the work by
Edalat et al.  \cite{edalat.97,potts.97a,potts.98,edalat.02}. There the authors
develop the general framework of representations using \emph{linear fractional
  transformations} that covers all known representations of real numbers that
are based on streams of nested intervals.  In particular the Edalat--Potts
\emph{normalisation algorithm} is a unified algorithm for computing all
elementary functions on real numbers.

The present work is part of the ongoing project of the author for formalising
and verifying the Edalat-Potts normalisation algorithm. We use the constructive
type theory extended with coinductive types to implement and formalise the
homographic and quadratic algorithms. These algorithms originated in the exact
continued fraction arithmetic~\cite{gosper.72,vuillemin.90,lester.01} and form
the basis of the Edalat--Potts algorithm.  The two algorithms suffice for
equipping the stream representations of real numbers with a field structure and
thus are important in themselves both from a theoretical and a practical point
of view.  The theoretical importance is highlighted when we consider our work as
a solution to the problem of equipping the \emph{coalgebraic} structure of real
numbers with the \emph{algebraic} properties of a field. This is due to the
innate relationship between coinductive types and final coalgebras which we
mention in Section~\ref{sec:coind-type}.

We use the machinery of the \coq proof assistant for coinductive types to
present the formalisation.  Throughout the article we use a syntax loosely based
on the \coq syntax, adapted for presenting in an article. We present definitions
and lemmas, depending on the usage context, as a \coq declaration (bound between
two horizontal bars) or as ordinary mathematical expressions.  Regarding lemmas
we follow this convention: If a lemma is to be used as a \coq subterm of another
lemma or definition then we will present its statement as a \coq declaration so
that it gets a reusable name; otherwise we present it as an ordinary numbered
lemma.  In order to improve the legibility of the \coq declarations we use the
curried version of the functions within \coq declarations, \eg \verb|F a b| will
be used in \coq mode and $F(a,b)$ within the text. All the lemmas in this
article are formalised and proven in \coq, a list of their machine checked
counterparts is given in Appendix~\ref{appendix:names_formalised}.  Most of the
proofs of lemmas are omitted in the paper, however in some cases the proof is
given in human language. In any case all the machine checked version of the
statements and proofs of all the lemmas and theorems can be found in the
complete \coq formalisation of the material in this article which is available
for download in~\cite{milad.07}.

%%%%%%%%%%%%%%%%%%%%%%%%%%%%%%%%%%%%%%%%%%%%%%%%%%%%%%%%%%%%%%%%%%%%%%%%%%%%%%%%
%                                                                              %
%                                Related Work                                  %
%                                                                              %
%%%%%%%%%%%%%%%%%%%%%%%%%%%%%%%%%%%%%%%%%%%%%%%%%%%%%%%%%%%%%%%%%%%%%%%%%%%%%%%%
\subsection*{Related Work}
The stream representation of exact real numbers have been recently formalised in
a coinductive setting by Ciaffaglione and Di
Gianantonio~\cite{ciaffaglione.06,ciaffaglione.03}, Bertot~\cite{bertot.06},
Hou~\cite{hou.06} and Gibbons~\cite{gibbons.06}.  Ciaffaglione and Di
Gianantonio use the \coq proof assistant to formalise a representation of real
numbers in $[-1,1]$ as ternary streams and to prove that paired with the natural
number exponents they form a complete Archimedean ordered field.  Bertot ---
using \coq as well--- formalises a ternary representation of $[0,1]$ using
\emph{affine maps} and formalises affine operations (multiplying by scalars),
addition, multiplication and infinite sums.  Hou studies two coinductive
representations of signed ternary digits and Cauchy sequences considered as
streams, proves their equivalence using set-theoretic coinduction and defines
the addition via the average function. Gibbons, as an application of his notion
of metamorphism, shows how one can transform various stream representations of
real numbers and use the same algorithms for different representations.

Our work, while related, is different from all of the above, in that we
formalise two powerful algorithms that give us all field operations on real
numbers, including division which seems to be the most difficult one in the
other approaches.  Furthermore due to the expressiveness of the Edalat--Potts
framework, the algorithms that we formalise are in principle independent of any
specific representation.  For presentation purposes we use a specific
representation, but our correctness proof can be adapted for other
representations. This is because the correctness proofs have several layers and
only one aspect of them is dependent upon the metric properties of the used
representation.  The coinductive aspect of our work is related to the above
work. For example we follow Bertot's and Hou's idea of using a coinductive
predicate to link real numbers and the streams representing
them~\cite{bertot.05a,bertot.06,hou.06}.  From a type theoretic point of view
the notion of \emph{cofixed point} equations has a central \role in our
development distinguishing it from the above work.

From a historical perspective, the Edalat--Potts algorithm was a step in
designing a programming language with a built-in abstract data-type for real
numbers~\cite{potts.97a}, in line with the work by Di
Gianantonio~\cite{digianantonio.93} and Escard\'{o}~\cite{escardo.96}. The
trade-off between the expressibility and the existence of parallelism in these
work led to some improvements on the domain theoretic semantics of the
Edalat--Potts algorithm, \eg as in the recent work in~\cite{marcial.07} where a
sequential language with a non-deterministic cotransitivity test is proposed.
This line of research is an instance of the \emph{extensional} approach to exact
real arithmetic while our work in which we have direct access to the digits of
the representation is a study in the \emph{intensional} exact arithmetic in the
sense of~\cite{bauer.02a}. However, the actual programs written in the
extensional approach do have a coalgebraic nature and are essentially
formalisable in the coinductive type theory.

In other related work, \pavlovic and Pratt~\cite{pavlovic.99} study the order
properties of the continuum as the final coalgebra for the list functor and
stream functor in category \Sets by specifying Cantor space and Baire space in
terms of these functors. However, by characterising the continuum only up to its
order type, their construction does not address the algebraic properties of real
numbers.  Freyd gives another characterisation of Dedekind reals
(see~\cite[\S~D4.7]{johnstone.02}) in terms of the diagonalisation of a `wedge'
functor in a category of posets.  In~\cite{escardo.01} the unit interval is
constructed as an initial algebra and the Cauchy completeness is defined by
uniqueness of a morphism from a coalgebra to an algebra.  The big picture that
we are working on, \ie the formalisation of the Edalat--Potts normalisation
algorithm is related to the work in \cite{pavlovic.98,simpson.98,pattinson.03}
that reconcile the coalgebraic structure of real numbers with algebraic
operations on them.

The general issue of formalising functions from streams to streams within
logical frameworks is studied by~\cite{mendler.86,paulson.94} (using
Knaster--Tarski's fixed point theorem) and \cite{matthews.00,digianantonio.03}
(using Banach--Mazur's fixed point theorem). Finally, the recent work
in~\cite{ghani.06} tackles this problem by internalising the notion of
\emph{productivity} in a single data type for all such functions. Productive
functions are those functions on infinite objects that produce provably infinite
output. The above formalisations all focus on formalising \emph{total}
productive functions. This is not surprising, given that in type theory we deal
with total functions. However, from the programmer's point of view, it might be
desirable to have a way of dealing with partial functions. Our work differs from
the above in that we embark on formalising partial algorithms on infinite
objects. In this sense our work is related to the work on formalising general
recursion for partial functions on inductive types~\cite{dubois.98,bove.01}.

Finally, our focus on the partial productivity is related to the aforesaid
domain-theoretic semantics~\cite{escardo.96} where partial real numbers are
denoted by interval and the \emph{strong convergence} (akin to our notion of
productivity) of the functions is studied~\cite{marcial.07}. This relationship
is not a coincidence as manifested by the original analytic proof of adequacy
for the Edalat--Potts algorithm~\cite{potts.97a}.

%%%%%%%%%%%%%%%%%%%%%%%%%%%%%%%%%%%%%%%%%%%%%%%%%%%%%%%%%%%%%%%%%%%%%%%%%%%%%%%%
%                                                                              %
%                        Type Theoretic Coinduction                            %
%                                                                              %
%%%%%%%%%%%%%%%%%%%%%%%%%%%%%%%%%%%%%%%%%%%%%%%%%%%%%%%%%%%%%%%%%%%%%%%%%%%%%%%%
\section{Type Theoretic Coinduction}\label{sec:coind-type}
The \coq proof assistant~\cite{coq} is an implementation of Calculus of
Inductive Constructions (\CIC) extended with coinductive types. This is is an
extension of \martinlof intensional type theory.  Coinductive types are intended
for accommodating infinite objects such as streams and infinite trees in type
theory~\cite{mendler.86}. This is in contrast to inductive types which are
accommodating well-founded and thus essentially finitistic objects such as
natural numbers and trees.  The coinductive types were added to \coq by
\gimenez~\cite{gimenez.96}.  Their implementation follows the same philosophy as
that of inductive types in \CIC, namely there is a general scheme that allows
for formation of coinductive types if their \emph{constructors} are given, and
if these constructors satisfy the \emph{strict positivity} condition.  The
definition of a strictly positive constructor is identical for inductive and
coinductive types and similar to that of a monomial endofunctor (\ie an
endofunctor involving products and exponentials).  Intuitively a constructor $c$
is strictly positive with respect to $x$ only if $x$ does not appear to the left
of a \arr in a nested occurrence of \arr in the type of $c$. A formal definition
can be found in~\cite{paulin.92}. This means that the following forms an
inductive (\resp coinductive) type $I$ in \coq, provided that the keyword
\verb|Inductive| (\resp \verb|CoInductive|) is given and that all $c_i$'s are
strictly positive constructors with respect to $I$.

\begin{coqtt}
(Co)Inductive \(I\) (\(x\sb{1}\)\(\colon\)\(X\sb{1}\))\dots(\(x\sb{i}\)\(\colon\)\(X\sb{i}\))\(\colon\)\(\forall\)(\(y\sb{1}\)\(\colon\)\(Y\sb{1}\))\dots(\(y\sb{m}\)\(\colon\)\(Y\sb{m}\))\(,\) \(s\):=
|\(c\sb{1}\)\(\colon\)\(\forall\)(\(z\sb{11}\)\(\colon\)\(Z\sb{11}\))\dots(\(z\sb{1k\sb{1}}\)\(\colon\)\(Z\sb{1k\sb{1}}\))\(,\) \(I\) \(t\sb{11}\) \dots \(t\sb{1{m{+}i}}\)
        \vdots
|\(c\sb{n}\)\(\colon\)\(\forall\)(\(z\sb{n1}\)\(\colon\)\(Z\sb{n1}\))\dots(\(z\sb{nk\sb{n}}\)\(\colon\)\(Z\sb{nk\sb{n}}\))\(,\) \(I\) \(t\sb{11}\) \dots \(t\sb{n{m{+}i}}\).
\end{coqtt}

In such a declaration $s$ is a sort,\ie $s\in\set{\Setcoq,\Propcoq,\Typecoq}$.
Moreover $x_{i}$s (\resp $y_i$s) are \emph{general} (\resp \emph{recursive})
parameters of $I$. 

For example one can define the set of streams as

\begin{coqtt}
CoInductive \Streams (A \(\colon\) \Setcoq) \(\colon\) \Setcoq :=
| \textcons \(\colon\) A \arr \Streams A \arr \Streams A.
\end{coqtt}
Note that this is a polymorphic type forming the streams of elements of its
general parameter $A$. From now on we shall use \(A^{\omega}\) to denote the
type \texttt{\Streams A}.

After a coinductive type is defined one can introduce its inhabitants and
functions into it. Such definitions are given by a \emph{cofixed point}
operator. This operator is similar to the fixed point operator for structural
recursion.  When given a well-typed definition that satisfies a
\emph{guardedness condition}, this operator will introduce an infinite object
that inhabits the coinductive type.

The typing rule for this operator is given by the following judgement (here, let
$I$ be a coinductive type with parameters $P_0,\dots,P_i$).
\begin{displaymath}
  \judgement
  {\begin{array}{l}
      \Gamma, f\colon  B \vdash N \colon B\\
      B \equiv \forall x_0\colon X_0,\dots,x_j\colon X_j,(I\appl P_0\appl\dots\appl P_i)\quad \textsf{G}(f,B,N)
    \end{array}}
  {\Gamma\vdash\cofix f \colon B := N \colon B}
\leqno{\cofix rule}
\end{displaymath}
           
According to this rule, if $f,B$ and $N$ satisfy the side condition \textsf{G}
then $\cofix f$ is an inhabitant of type $B$ which is a function type with as
codomain a coinductively defined type. In this case $N$ is the body of the
definition which may contain $f$.  The side condition $\textsf{G}(f,B,N)$ is
called the \coq guardedness condition and is a syntactic criterion that is
intended to ensure the productivity of infinite objects.  This condition checks
whether the declaration of $f$ is guarded by constructors. This means that every
occurrence of $f$ in the body $N$ should be the immediate argument of a
constructor of some inductive or coinductive type. Note that it need not be the
argument of only the constructors of $I$, and that the constructors can
accumulate on top of each other. Thus $f$ occurs guarded if it occurs as $c_0
(c_1\dots (c_m\appl f)\dots)$ where each $c_i$ is a constructor of some
inductive or coinductive type $I_i$. This condition is due to
\gimenez~\cite{gimenez.96} and is based on earlier work of
Coquand~\cite{coquand.94}.  A precise definition of \textsf{G} can be found
in~\cite[p.~175]{gimenez.96}.

Finally we mention the reduction (in fact expansion) rule corresponding to the
\cofix operator. Let $F \equiv \cofix g\colon B := N$. Then the \emph{cofixed
  point expansion} is the following rule.
\begin{multline*}
\textbf{match } (F\appl P_0\dots P_j)\colon X \textbf{ with }  \mathrel{|} r_0
      \Rightarrow R_0 \mathrel{|} \dots \mathrel{|} r_k \Rightarrow
      R_k \textbf{ end}\leadsto\\  \textbf{match }
      (N[g\gets F]\appl P_0\dots P_j)\colon X \textbf{ with }  \mathrel{|} r_0
      \Rightarrow R_0 \mathrel{|} \dots \mathrel{|} r_k \Rightarrow
      R_k \textbf{ end}\enspace.
\end{multline*}
Thus, the expansion of a cofixed point is only allowed when a case analysis of
the cofixed point is done.

It is well-known that coinductive types correspond to weakly final coalgebras in
categorical models of intensional type theory~\cite{hagino.87}.  From a
coalgebraic point of view this treatment of coinductive types by means of
constructors and cofixed point operator might seem unnatural: final coalgebras
are about observations and not constructions; final coalgebra should be given
using its destructor.  Nevertheless, presenting the coinductive types in the
\coq way, is much closer to the syntax of lazy functional programming languages
such as \haskell\footnote{Note that in \haskell --- where there is no
  distinction between inductive and coinductive types--- all data-types can be
  considered to be potentially infinite and hence correspond to \coq's
  coinductive types.} and hence very useful for many applications. Moreover, as
we show in Section~\ref{sec:general-corec}, one can use \coq to define a general
form of productive functions, allowing one to build more complicated coalgebraic
structures.  In any case, theoretically this does not change the coalgebraic
semantics and the coinductive types can still be interpreted as weakly final
coalgebras in any categorical model of \CIC (see~\cite{abbott.05} where a
stronger results is proven). Furthermore, the usual coiteration and corecursion
schemes can be derived in terms of the operator \cofix~\cite{gimenez.95}.
Therefore the method that we present in this article using the language of \coq
can easily be translated into the standard categorical notations in any
categorical model of \CIC\footnote{In fact, in the present article we do not
  need the \emph{universes} in \CIC and therefore categorical models of simpler
  extensions of \martinlof type theory --- such as \martinlof categories of
  Abbott et al~\cite{abbott.05} --- will suffice.}.

The guardedness condition of \coq is one among many syntactic criteria for
ensuring productivity. Examples include \emph{corecursion}~\cite{geuvers.92},
\emph{dual of course of value recursion}~\cite{uustalu.99},
\emph{$T$-coiteration for pointed functors}~\cite{lenisa.99},
\emph{$\lambda$-coiteration for distributive laws}~\cite{bartels.01} and
\emph{bialgebraic $T$-coiteration} \cite{cancila.03}, each handling an ever
expanding class of specifications.  However, the productivity of the algorithms
on real numbers cannot be syntactically detected. In fact the productivity of
the standard filter function on stream of natural numbers with the following
specification is also not decidable (here $P$ is a boolean predicate on natural
numbers and we use $x\cons xs$ to denote $\textcons\appl x\appl xs$).
\begin{displaymath}
  \filter\appl  (x\cons xs) :=
  \begin{cases}
    x \cons\appl \filter xs &\qquad \ppif P(x)\enspace, \\
    \filter\appl  xs &\qquad         \ppow
  \end{cases}
\end{displaymath}
By suitably choosing $P$ one can reduce the problem of the productivity of the
above function to an open problem in mathematics;
see~\cite[Example~4.7.6]{milad.04b} for a choice of $P$ which shows that the
productivity of the above function is equivalent to whether there are infinitely
many twin prime numbers.

Therefore it seems that providing syntactic productivity tests cannot cover the
most general class of recursive specifications for infinite objects.  One
possible solution is to adhere to semantic means in order to be able to
formalise such programs using one of the above schemes. For instance, for the
case of filter on prime numbers, one has to (1) consider a number theoretic
constructive proof of the infinitude of primes, (2) from this proof extract a
function $\kappa$ that returns the $n$th prime number, (3) use $\kappa$ to
rewrite \filter in a way that it passes syntactic tests of productivity, \ie
using one of the above syntactic schemes~\cite[\S~4.7]{milad.06c,milad.04b}.

Another work-around, one that we follow in this article, is to adhere to
advanced type-theoretic methods to bypass this condition. This is similar to the
application of dependent inductive types for formalising general recursion using
structural recursion~\cite{dubois.98,bove.01}. For coinductive types this has
led to a method of general corecursion for filter-like
functions~\cite{bertot.05} and a similar method that we use in
Section~\ref{sec:general-corec} for formalising the homographic and quadratic
algorithms in \coq.

The \cofix operator and its expansion rule together with the guardedness
condition constitute the machinery of the \coq system for coinductive types.
This means that there is no separate tool for proofs by \emph{coinduction}. This
is in contrast to the set-theoretic greatest fixed point semantics for
coinduction where for each coinductive object a coinduction proof principle is
present which is inherent in the monotonicity of the set operator
\cite{barwise.96}. Instead in the type theoretic approach, where proofs are
objects too, we use the \cofix operator to directly \emph{build} the coinductive
proof as a proof object. This means that whenever we want to prove by
coinduction, our goal should be a coinductive type. If necessary, specialised
coinductive predicates should be created for formalising a proof that uses
coinduction.  These additional predicates are in most cases straightforward
reformulation of the corresponding set-theoretic proof principle (\cf the
extensional equality \bisim below). However, sometimes special care has to be
taken to overcome the restrictions put forward by the guardedness condition (\cf
\rep in Section~\ref{sec:representation}). As a result, \coq's direct approach
to coinduction makes the coinductive proofs easier than their set-theoretic
counterparts as long the guardedness condition does not get in the way.

For proving equalities by coinduction, in coalgebraic and set-theoretic settings
one relies on the notion of \emph{bisimulation}~\cite{barwise.96,jacobs.97}. In
the case of streams, a bisimulation is a binary relation $R$ satisfying the
property that
\begin{displaymath}
  R(\alpha,\beta) \implies \hd(\alpha)=\hd(\beta) \wedge R(\tl(\alpha),\tl(\beta))\enspace.
\end{displaymath}
Here \hd and \tl are functions on streams that give the head and the tail of a
stream.  Then one can prove that two streams are equal if they satisfy a
bisimulation relation.  The coinduction proof principle thus consists of finding
a suitable bisimulation.

To translate this proof principle into the type theoretic coinduction note that
the bisimulation relation leads to the extensional equality, which in the
\emph{intensional} type theories, such as \CIC, is quite distinct from the
built-in notion of equality. In fact each extensional equality should be defined
and added to the type system. On the other hand, recall that we can only prove
by coinduction in \coq if the goal of the proof has a coinductive type. This
leads us to the following definition for a coinductive extensional equality on
streams which we denote by \bisim.

\begin{coqtt}
CoInductive \(\bisim\) \(\colon\) \(A\sp{\omega}\) \arr \(A\sp{\omega}\) \arr\Propcoq :=
|  \(\bisim\sb{c}\) \(\colon\)\(\forall\)(\(\alpha\sb{1} \alpha\sb{2}\colon A\sp{\omega}\)), \(\hd \alpha\sb{1}=\hd \alpha\sb{2}\) \arr\(\tl \alpha\sb{1}\bisim \tl \alpha\sb{2}\) \arr\(\alpha\sb{1}\bisim\alpha\sb{2}\).
\end{coqtt}
Note that \(\bisim\sb{c}\), the sole constructor of \bisim, has the shape of a
bisimulation property. The proof that this is an equivalence relation can be
found in the standard library of \coq~\cite{coq}.  Moreover, \gimenez shows that
this is a bisimulation equivalence relation and derives the usual principle of
coinduction~\cite[\S~4.2]{gimenez.96}.  In the present work we use \bisim
relation in our coinductive correctness proofs.

%%%%%%%%%%%%%%%%%%%%%%%%%%%%%%%%%%%%%%%%%%%%%%%%%%%%%%%%%%%%%%%%%%%%%%%%%%%%%%%%
%                                                                              %
%                     Homographic and Quadratic Algorithms                     %
%                                                                              %
%%%%%%%%%%%%%%%%%%%%%%%%%%%%%%%%%%%%%%%%%%%%%%%%%%%%%%%%%%%%%%%%%%%%%%%%%%%%%%%%
\section{Homographic and Quadratic Algorithms}\label{sec:algorithms}
The homographic and quadratic algorithms are similar to Gosper's
algorithm~\cite{gosper.72} for addition and multiplication on continued
fractions and form the basis of the Edalat--Potts approach to lazy exact real
arithmetic~\cite{edalat.97,potts.98}.

Here we use a representation which is much simpler than the continued fractions
but it is redundant enough to ensure productivity\footnote{The necessity of
  redundancy in the representations for real numbers is studied in the framework
  of computable analysis~\cite{weihrauch.00} but is it outside the scope of the
  present article.}.  There is nothing special about this representation apart
from the fact that it eases the \coq formalisation of the proofs of the metric
properties that we use in this work, thus giving us a prototype formalisation of
the algorithms that is concrete and hence can be computed with.  A treatment of
the general case where we abstract away both the digit set and the compact
subinterval of $[-\infty,+\infty]$ can be found in~\cite[\S~5]{milad.04b}.
Thus, for practical and presentational purposes, we consider a fixed
representation for \Ibase containing $3$ digits, each of which a \mobius map.
\emph{\mobius maps} are maps of the form
\begin{displaymath}
  x\longmapsto \frac{ax+b}{cx+d}\enspace,
\end{displaymath} 
where\footnote{Note that we could equivalently take the coefficients to be in
  \ZZ.} $a,b,c,d\in\QQ$. \mobius maps are usually denoted by the matrix of their
coefficients.  A \mobius map is \emph{bounded} if its denominator does not
vanish in \Ibase.  A \mobius map is \emph{refining} if it maps the closed
interval \Ibase into itself. Assuming $\mu={\slft a b c d}$ we introduce two
predicates that capture these properties:
\begin{align*}
\bounded{\mu} &:= 0{<}d{+}c \wedge 0{<}d{-}c \bigvee d{+}c{<}0 \wedge d{-}c{<}0\enspace,\\
\refining{\mu} &:= \bounded{\mu} \bigwedge\\
              &\quad\quad (0{<}a{+}b{+}c{+}d \wedge 0{<}a{-}b{-}c{+}d \wedge 0{<}{-}a{-}b{+}c{+}d \wedge 0{<}{-}a{+}b{-}c{+}d \bigvee\\
              &\quad\qquad a{+}b{+}c{+}d{<}0 \wedge a{-}b{-}c{+}d{<}0 \wedge {-}a{-}b{+}c{+}d{<}0 \wedge {-}a{+}b{-}c{+}d{<}0)\enspace.
\end{align*}

For our representation, we consider the set $\DIG=\set{\LL,\RR,\MM}$ and denote
the set of streams of elements of \DIG by \DIGinf. We interpret each digit by a
refining \mobius map as follows.
\begin{displaymath}
 \LL=\slft{\frac{1}{2}}{\frac{-1}{2}}{\frac{1}{2}}{\frac{3}{2}}\enspace,\quad\RR=\slft{\frac{1}{2}}{\frac{1}{2}}{\frac{-1}{2}}{\frac{3}{2}}\enspace,\quad\MM=\slft{1}{0}{0}{3}\enspace.
\end{displaymath}
In fact under the conjugacy map $S(x)=\frac{x-1}{x+1}$, these are the conjugates
of the Stern--Brocot representation for $[0,+\infty]$ presented in
~\cite[\S~5.7]{milad.04b}, hence the fact that \DIGinf is a representation for
\Ibase is easily derivable form the properties of the Stern--Brocot
representation (see also Section~\ref{sec:representation}).

The \emph{homographic algorithm} is the algorithm that given a \mobius map $\mu$
and a stream $\alpha\in\DIGinf$ representing $r_{\alpha}$, outputs a stream
$\gamma$ that represents $r_{\gamma}$ such that $\mu(r_{\alpha})=r_{\gamma}$.
In order to present the homographic algorithm we need an \emph{emission
  condition} \emitting{\mu}{\phi} for a digit $\phi={\slft {\phi_{00}}
  {\phi_{01}} {\phi_{10}} {\phi_{11}}}$ and $\mu$ which checks the
inclusion of intervals $\mu(\Ibase) \subseteq \phi(\Ibase)$.
\begin{align*}
\emitting{\mu}{\phi} := \bounded{\mu} \wedge & (d{-}c)(d{-}c)(\phi_{01}{-}\phi_{00})\leq (d{-}c)(b{-}a)(\phi_{11}{-}\phi_{10}) \wedge\\  
             &(d{-}c)(b{-}a)(\phi_{10}{+}\phi_{11})\leq (d{-}c)(d{-}c)(\phi_{00}{+}\phi_{01}) \wedge\\
             &(d{+}c)(c{+}d)(\phi_{01}{-}\phi_{00})\leq (d{+}c)(a{+}b)(\phi_{11}{-}\phi_{10}) \wedge \\
             & (d{+}c)(a{+}b)(\phi_{10}{+}\phi_{11})\leq (d{+}c)(c{+}d)(\phi_{00}{+}\phi_{01})\enspace. 
\end{align*}

Note that since the above are expressions involving only rational numbers the
emission condition is a decidable predicate.  This enables us to state the
homographic algorithm:
\begin{multline*}
  \homographic\appl \mu\appl (x\cons xs) :=
  \\
  \begin{cases}
     \LL \cons \homographic\appl (\LL^{-1} \after \mu)\appl (x\cons xs)&\qquad \ppif \emitting{\mu}{\LL}\enspace,\\
     \RR \cons \homographic\appl (\RR^{-1} \after \mu)\appl (x\cons xs)&\qquad \ppelseif \emitting{\mu}{\RR}\enspace,\\
     \MM \cons \homographic\appl (\MM^{-1} \after \mu)\appl (x\cons xs)&\qquad \ppelseif \emitting{\mu}{\MM}\enspace,\\
     \homographic\appl\mu \after x\appl xs&\qquad\ppow\\
  \end{cases}
\end{multline*}
Here $d^{-1}$ and $\after$ denote the usual matrix inversion and matrix product.
The first three branches (\resp the last branch) are called \emph{emission
  steps} (\resp \emph{absorption step}). Note that due to the redundancy of the
representation, the case distinction need not be mutually exclusive, but this
does not affect the outcome.

The intuition behind the algorithm is that we start by considering an infinite
product of \mobius maps, of which all but the first one are digits. We start
pushing $\mu$ towards the infinity by absorbing digits (hence obtaining a new
refining \mobius map) and emitting digits whenever the emission condition holds,
\ie whenever the range of \mobius map applied to the interval \Ibase fits inside
the range of a digit.
\begin{displaymath}
\mu\after \phi_0\after \phi_1 \after\cdots \qquad\leadsto\qquad \phi\after (\phi^{-1}\after\mu) \after \phi_0\after \phi_1\after\cdots \quad\ppif \emitting{\mu}{\phi}\enspace.
\end{displaymath}
For a more formal semantics for the algorithm see~\cite{potts.97a} and the
semantical proof of correctness that is given in~\cite[\S~5.6]{milad.04b}.

To compute binary algebraic operations we consider the \emph{quadratic map}
which is a map
\begin{displaymath}
  \xi(x,y):= \frac{axy+bx+cy+d}{exy+fx+gy+h}\enspace,
\end{displaymath}
with $a,b,c,d,e,f,g\in \QQ$ and can be denoted by its $2\times 2\times 2$ tensor
of coefficients.  A quadratic map is \emph{bounded} if its denominator does not
vanish in $\Ibase\times\Ibase$.  A \emph{refining} quadratic map is a quadratic
map $\xi$ such that $\xi(\Ibase,\Ibase)\subseteq \Ibase$. The predicates
$\bounded{\xi}$ and $\refining{\xi}$ can easily be stated in terms of
inequalities on rational numbers (see
Appendix~\ref{appendix:quadratic-predicates}).

The \emph{quadratic algorithm} is an algorithm that given a quadratic map $\xi$
and two streams $\alpha,\beta\in\DIGinf$ representing $r_{\alpha}$ and
$r_{\beta}$, outputs a stream $\gamma$ that represents $r_{\gamma}$ such that
$\xi(r_{\alpha},r_{\beta})=r_{\gamma}$. Here too we need a decidable emission
condition \emitting{\xi}{\phi} that checks the inclusion of intervals
$\xi(\Ibase,\Ibase) \subseteq \phi(\Ibase)$ for each digit $\phi$; its explicit
definition is given in Appendix~\ref{appendix:quadratic-predicates}. By
$\mu\afterM\xi$ we denote the composition of a \mobius map $\mu$ and a
quadratic map $\xi$ (note that the outcome is again a quadratic map).  Moreover
we use $\xi\afterL \mu$ and $\xi\afterR \mu$ to denote the two different ways
of composing a quadratic map and a \mobius map by considering the \mobius map as
its first (\resp second) argument. With this notation we can present the
quadratic algorithm:
\begin{multline*}
  \quadratic\appl \xi\appl (x\cons xs)\appl (y\cons ys):=
  \\
  \begin{cases}
     \LL \cons \quadratic\appl (\LL^{-1} \afterM \xi)\appl (x\cons xs)\appl(y\cons ys)
                                                                       &\qquad \ppif \emitting{\xi}{\LL}\enspace,\\
     \RR \cons \quadratic\appl (\RR^{-1} \afterM \xi)\appl (x\cons xs)\appl(y\cons ys)
                                                                       &\qquad \ppelseif \emitting{\xi}{\RR}\enspace,\\
     \MM \cons \quadratic\appl (\MM^{-1} \afterM \xi)\appl (x\cons xs)\appl(y\cons ys)
                                                                       &\qquad \ppelseif \emitting{\xi}{\MM}\enspace,\\
     \quadratic\appl(\xi\afterL x\afterR y)\appl xs\appl ys&\qquad\ppow\\
  \end{cases}
\end{multline*}
The intuition behind this algorithm is similar to the homographic algorithm.
The homographic algorithm can be used to compute the unary field operation of
opposite, while the quadratic algorithm can be used for binary field operations
of addition, multiplication and division. Simply taking $\xi:={\stensor 1 0 0 0
  0 0 0 1}$ it gives the multiplication. Note that addition and division are not
total functions on \Ibase. The quadratic algorithm applied to $\xi:={\stensor 0
  1 1 0 0 0 0 1}$ is a partial function that will calculate the addition (and
also it is productive) if and only if the inputs add up to a number within
\Ibase. However, the algorithms can also be used to calculate the binary
\emph{average} function and the restricted division that are defined
in~\cite{ciaffaglione.06}.

In the present work we do not study the total version of field operations and
computations on the whole real line.  However, we mention that transferring the
computation to the whole real line is possible.  A possibility would be to first
move to $[0,+\infty]$ via the inverse of the above conjugacy map. Form here we
can follow~\cite{potts.98} where a redundant sign bit is added by considering a
fourth order elliptic \mobius map that leads to a cyclic group consisting of
four signs for an unbiased exact floating point~\cite[\S 9.2]{potts.98}.

%%%%%%%%%%%%%%%%%%%%%%%%%%%%%%%%%%%%%%%%%%%%%%%%%%%%%%%%%%%%%%%%%%%%%%%%%%%%%%%%
%                                                                              %
%                   Homographic and Quadratic General Corecursive              %
%                                                                              %
%%%%%%%%%%%%%%%%%%%%%%%%%%%%%%%%%%%%%%%%%%%%%%%%%%%%%%%%%%%%%%%%%%%%%%%%%%%%%%%%
\section{General Corecursive Version of the algorithms}\label{sec:general-corec}
Algorithms of the previous section specify partial functions into the
coinductive type of streams. This partiality is problematic for us.  Translating
these specifications into the language of \coq means that we should ensure that
the returned value is provably an infinite stream, which is obviously not always
true for a partial function.  The algorithms resemble the general shape of the
\filter algorithm (see Section~\ref{sec:coind-type}).  Hence, as expected, they
do not satisfy the guardedness test of \coq, and indeed any other one of the
syntactic schemes used in the theory of coalgebras.

The usual way of dealing with partial functions in type theory is to consider
them as total functions but on a new, restricted domain which corresponds to the
values on which the partial function is defined.  In our case, it is
well-known~\cite{potts.98} that the algorithms are productive if they are
applied with refining maps. Proof of this fact is a tedious semantic proof that
deals with rational intervals.  Thus we need to incorporate among the arguments
of the function an additional argument, a so called \emph{proof obligation},
that captures the property of being refining and hence the semantic proof of the
infinitude of the outcome.  But directly adding the refining property
\refiningvoid, does not give us enough type theoretic machinery because
\refiningvoid is just a simple propositional predicate that lacks any inductive
or coinductive structure.

Instead our proposed proof obligation will have a more complex shape, enabling
us to use type theoretic tools of structural recursion and coinduction. Later in
Section~\ref{sec:topologic-correctness} we show that our proposed proof
obligation is a consequence of \refiningvoid. Instead of relying on properties
of interval inclusion our predicate will rely on the intuitive idea of having an
infinite output. Such a proof obligation is satisfied if at every step in the
algorithm after absorbing a finite number of digits the emission condition
eventually holds and hence we can output a digit.  We plan to capture this
inside a recursive function that at each step outputs the next digit, serving as
a modulus for productivity. The original algorithms will then call this function
at every step to obtain the next digit while keeping track of the new arguments
that should be passed to future step. This idea is used by
Bertot~\cite{bertot.05} to give a general method for defining \filter in \coq.
In this section we apply a modification of Bertot's method for our algorithms of
exact arithmetic.

%%%%%%%%%%%%%%%%%%%%%%%%%%%%%%%%%%%%%%%%%%%%%%%%%%%%%%%%%%%%%%%%%%%%%%%%%%%%%%%%
%                                                                              %
%                        Homographic General Corecursive                       %
%                                                                              %
%%%%%%%%%%%%%%%%%%%%%%%%%%%%%%%%%%%%%%%%%%%%%%%%%%%%%%%%%%%%%%%%%%%%%%%%%%%%%%%%
\subsection{Homographic Algorithm} \label{subsec:general-corec-homographic} Let
\MAT (\resp \TEN) be the set of \mobius maps (\resp quadratic maps) in
\coq\footnote{They can be considered as $\QQ^{4}$ and $\QQ^{8}$ respectively,
  forgetting about the refining and nonsingular properties. Those properties
  will enter the picture when we study the correctness of the algorithms.}.  We
are seeking to define a map $\map{h}{\MAT \times \DIGinf}{\DIGinf}$, that
corresponds to the homographic algorithm.  But $h$ is a partial function and is
not productive at every point. So instead of defining $h$ we shall define a map
\begin{displaymath}
\map{\bar{h}}{\Pi (\mu\colon \MAT)(\alpha\colon\DIGinf).
  P_h(\mu,\alpha)}{\DIGinf}
\end{displaymath}
where $P_h(\mu,\alpha)$ is a predicate (\ie a term of type \Propcoq) with the
intended meaning that the specification of the homographic algorithm is
productive when applied to $\mu$ and $\alpha$. In other words it specifies the
domain of the partial function $h$. We shall call $P\sb{h}$ a \emph{productivity
  predicate}.

The definition of $P\sb{h}$ is based on the modulus of productivity. This
modulus is a recursive function
\begin{displaymath}
\map{m_h}{\MAT\times
  \DIGinf}{\DIG\times\MAT\times\DIGinf}
\end{displaymath}
with the intended meaning that
$m_h(\mu,\alpha)=\pair{\phi}{\pair{\mu'}{\alpha'}}$ if and only if
\begin{displaymath}
  \homographic\appl \mu\appl\alpha \qquad\rightsquigarrow\qquad \phi\cons \homographic\appl \mu'\appl\alpha'\enspace,
\end{displaymath}
where `$\rightsquigarrow$' denotes multiple reduction steps after which $\phi$
is output (so after output of $\phi$ there are no more digits absorbed in
$\mu'$).  We would like this to be a function with recursive calls on $\alpha$,
but this is not possible. The reason is that $\alpha$ has a coinductive type
while in the structural recursion scheme we need an element with an inductive
type.  In other words we need to accommodate the domain of the function $m_h$
with an inductively defined argument which will be used for recursive calls.

This situation is similar to the case of partial recursive functions or
recursive functions with non-structurally recursive arguments. In order to
formalise such function in constructive type theory, there is a method of adding
an inductive domain predicate introduced in~\cite{dubois.98} and extensively
developed by Bove and Capretta~\cite{bove.01}.  According to this method we need
to define an inductively defined predicate $E_h(\mu,\alpha)$ with the intended
meaning that $\mu$ and $\alpha$ are in the domain of $m_h$ which in turn means
that the homographic algorithm should emit at least one digit when applied on
$\mu$ and $\alpha$.  Thus, as a first step in the definition of the productivity
predicate, we define $E_h$ as the following inductive type.

\begin{coqtt}
Inductive \(E\sb{h}\)\(\colon\) \(\MAT\) \arr \(\DIGinf\) \arr \Propcoq :=
|\(E\sb{hL}\)\(\colon\)\(\forall\)(\(\mu\colon\MAT\))(\(\alpha\colon\DIGinf\)), \emitting{\mu}{\LL} \arr \(E\sb{h}\) \(\mu\) \(\alpha\)
|\(E\sb{hR}\)\(\colon\)\(\forall\)(\(\mu\colon\MAT\))(\(\alpha\colon\DIGinf\)), \emitting{\mu}{\RR} \arr \(E\sb{h}\) \(\mu\) \(\alpha\)
|\(E\sb{hM}\)\(\colon\)\(\forall\)(\(\mu\colon\MAT\))(\(\alpha\colon\DIGinf\)), \emitting{\mu}{\MM} \arr \(E\sb{h}\) \(\mu\) \(\alpha\)
|\(E\sb{hab}\)\(\colon\)\(\forall\)(\(\mu\colon\MAT\))(\(\alpha\colon\DIGinf\)), \(E\sb{h}\) (\(\mu\after(\hd \alpha)\)) (\(\tl\) \(\alpha\)) \arr \(E\sb{h}\) \(\mu\) \(\alpha\).
\end{coqtt}
Here $E\sb{hL},E\sb{hR},E\sb{hM}$ and $E\sb{hab}$ are constructors of $E\sb{h}$.
Note that \(E\sb{h}\) has one constructor for each branch of the homographic
algorithm.

This allows us to define the modulus of productivity, \ie a recursive function
\begin{displaymath}
  \map{\overline{m}_h}{\Pi (\mu\colon\MAT)(\alpha\colon\DIGinf).
    E\sb{h}(\mu,\alpha)}{\DIG\times\MAT\times\DIGinf}
\end{displaymath}
as follows.

\begin{coqtt}
Fixpoint \(\overline{m}\sb{h}\)(\(\mu\colon\MAT\))(\(\alpha\colon\DIGinf\))(\(t\colon\!E\sb{h} \mu \alpha\))\{struct \(t\)\}\(\colon\)\DIG\!\!\!*(\MAT\!\!\!*\DIGinf):=
match \emittingdec{\mu}{\LL} with
| left _\(\Rightarrow\) \pair{\LL}{\pair{\LL\sp{-1}\after\mu}{\alpha}}
| right \(t\sb{l}\)\(\Rightarrow\)
   match \emittingdec{\mu}{\RR} with
   | left _\(\Rightarrow\) \pair{\RR}{\pair{\RR\sp{-1}\after\mu}{\alpha}}
   | right \(t\sb{r}\)\(\Rightarrow\)
      match \emittingdec{\mu}{\MM} with
      | left _\(\Rightarrow\) \pair{\MM}{\pair{\MM\sp{-1}\after\mu}{\alpha}}
      | right \(t\sb{m}\)\(\Rightarrow\) \(\overline{m}\sb{h}\) (\(\mu\after(\hd \alpha)\)) (\(\tl\) \(\alpha\)) (\(E\sb{hab}\_\textrm{inv}\) \(\mu\) \(\alpha\) \(t\sb{l}\) \(t\sb{r}\) \(t\sb{m}\) \(t\))
      end
   end
end.
\end{coqtt}
Here \verb|Fixpoint| (\resp \verb|struct|) are \coq keywords to denote a
recursive definition (\resp recursive argument of structural recursive calls).
Moreover, in the body of the definition two terms \emittingdecvoid and
\(E\sb{hab}\_\textrm{inv}\) are used. Both terms can be proven as lemmas in
\coq. The first lemma is the following.

\begin{coqtt}
Lemma \emittingdecvoid\(\colon\)\(\forall\) (\(\mu\colon\MAT\)) (\(\phi\colon\DIG\)), \emitting{\mu}{\phi} \sumbool \(\neg\)\emitting{\mu}{\phi}.
\end{coqtt}
This term extracts the informative computational content of the predicate
\emittingvoid which is a term of the type \Propcoq. This is necessary because in
\CIC one cannot obtain elements of the type \Setcoq by pattern matching on
propositions. Thus we have to use
\map{\sumbool}{\Propcoq{\times}\Propcoq}{\Setcoq} --- with \verb|left| and
\verb|right| its coprojections--- to transfer propositions into a boolean sum on
which we can pattern match. Hence the need for the above lemma is inevitable,
although its proof is quite trivial.

The second lemma states an inverse of the last constructor of \(E\sb{hab}\) in
case no emission condition holds.

\begin{coqtt}
Lemma \(E\sb{hab}\_\textrm{inv}\)\(\colon\)\(\forall\)(\(\mu\colon\MAT\))(\(\alpha\colon\DIGinf\)),
      \(\neg\)\emitting{\mu}{\LL}\arr\(\neg\)\emitting{\mu}{\RR}\arr\(\neg\)\emitting{\mu}{\MM}\arr\(E\sb{h}\) \(\mu\) \(\alpha\) \arr
                                        \(E\sb{h}\) (\(\mu\after(\hd \alpha)\)) (\(\tl\) \(\alpha\)).
\end{coqtt}
This lemma can be proven because \(E\sb{h}\) is an inductive type and hence all
its canonical objects should be generated by one of its
constructors\footnote{Due to some technical issues with respect to the type
  theory of \coq, the proof has to be built using a specific method that is
  described in details in~\cite[\S 15.4]{bertot.04}. These issues are out of the
  scope of the present work.}

Note that in $\overline{m}_h$ the output is independent of the proof $t$. The
term $t$ only serves as a catalyst that allows for using recursion where all the
other arguments are not inductive. Thus we should be able to prove a \emph{proof
  irrelevance} result for \(\overline{m}\sb{h}\).

\begin{lem}\label{lemma:modulus_h_PI}
  Let $\mu\in\MAT,\alpha\in\DIGinf$ and $t_1,t_2$ be two proofs that
  $E\sb{h}(\mu,\alpha)$ holds. Then
  $$%\begin{displaymath}
    \overline{m}\sb{h}(\mu,\alpha,t\sb{1}) = \overline{m}\sb{h}(\mu,\alpha,t\sb{2})\enspace.\eqno{\qEd}
  $$%\end{displaymath}
\end{lem}

The proof of the above lemma is based on a dependent induction scheme for
\(E\sb{h}\) that is more specialised than the usual induction scheme attributed
to the inductive types: the ordinary induction scheme can be used to prove a
property $R\colon \MAT\arr\DIGinf\arr\Propcoq$ while the dependent induction
scheme can be used to prove a property
\begin{displaymath}
  R\colon
  \Pi(\mu\colon\MAT)(\alpha\colon\DIGinf).  E_{h}(\mu,\alpha)\arr\Propcoq\enspace.
\end{displaymath}

The Lemma~\ref{lemma:modulus_h_PI} enables us to prove the \emph{fixed point
  equations} of the $\overline{m}\sb{h}$ function. These are in fact unfolding
of the body of the definition of $\overline{m}\sb{h}$; they are crucial for
proving similar results for the homographic algorithm. Hence we mention them in
a lemma here:

\begin{lem} \label{lemma:modulus_h_fixedpoints}
  Let $\mu\in\MAT,\alpha\in\DIGinf$ and $t$ be a proof that $E\sb{h}(\mu,\alpha)$
  holds. 
  \begin{enumerate}[\em(1)]
  \item If \,\emitting{\mu}{\LL} holds then 
    \begin{displaymath}
      \overline{m}\sb{h}(\mu,\alpha,t) =
      \pair{\LL}{\pair{\LL\sp{-1}\after\mu}{\alpha}}\enspace.
    \end{displaymath} \label{item:modulus_h_fixedpoints_1}
  \item If \,\(\neg\emitting{\mu}{\LL}\) but \emitting{\mu}{\RR} holds then
    \begin{displaymath}
      \overline{m}\sb{h}(\mu,\alpha,t) =
      \pair{\RR}{\pair{\RR\sp{-1}\after\mu}{\alpha}}\enspace.
    \end{displaymath} \label{item:modulus_h_fixedpoints_2}
  \item If \,\(\neg\emitting{\mu}{\LL}\) and \(\neg\emitting{\mu}{\RR}\) but
    \emitting{\mu}{\MM} holds then
    \begin{displaymath}
      \overline{m}\sb{h}(\mu,\alpha,t) =
      \pair{\MM}{\pair{\MM\sp{-1}\after\mu}{\alpha}}\enspace.
    \end{displaymath} \label{item:modulus_h_fixedpoints_3}
  \item If \,\(\neg\emitting{\mu}{\LL}\), \(\neg\emitting{\mu}{\RR}\) and
    \(\neg\emitting{\mu}{\MM}\) holds then for all $t'$ a proof of property
    $E\sb{h}\big(\mu\after(\hd(\alpha)),\tl(\alpha)\big)$ we have
    $$%\begin{displaymath}
      \overline{m}\sb{h}(\mu,\alpha,t) = \overline{m}\sb{h}\big(\mu\after(\hd(\alpha)),\tl(\alpha),t'\big)\enspace.\eqno{\qEd}
    $$%\end{displaymath} 
    \label{item:modulus_h_fixedpoints_4}
  \end{enumerate}
\end{lem}

Note that the last part states a more general fact than just the fourth branch
of the recursive definition of $\overline{m}\sb{h}$ because the proof obligation
\(t'\) is abstracted. Nevertheless its proof is similar to the other three
parts.

Having defined $\overline{m}_h$ we need one more auxiliary predicate before
defining $P_h$. This auxiliary predicate is an inductive predicate that ensures
that $E_{h}$ holds for some finite iteration of $\overline{m}_h$ (here
$\pi_{ij}$ denotes the $i$-th projection of a $j$-tuple).

\begin{coqtt}
Inductive \(\Psi\sb{h}\)\(\colon\)\(\NN\)\arr\MAT\arr\DIGinf\arr\Propcoq :=
|\(\Psi\sb{h0}\)\(\colon\)\(\forall\)(\(\mu\colon\MAT\))(\(\alpha\colon\DIGinf\)), \(E\sb{h} \mu \alpha\) \arr \(\Psi\sb{h} 0 \mu \alpha\)
|\(\Psi\sb{hS}\)\(\colon\)\(\forall\)(\(n\colon\NN\))(\(\mu\colon\MAT\)) (\(\alpha\colon\DIGinf\)) (\(t\colon\!E\sb{h} \mu \alpha\)), 
       \(\Psi\sb{h}\) \(n\) (\(\pi\sb{23}\)\((\overline{m}\sb{h} \mu \alpha t)\)) (\(\pi\sb{33}\)\((\overline{m}\sb{h} \mu \alpha t)\)) \arr\(\Psi\sb{h}\) \((n{+}1)\) \(\mu\) \(\alpha\).
\end{coqtt}

We use the above predicate to define $P_h$, a predicate that captures the
productivity of the homographic algorithm. This predicate will be an inductive
type with one constructor.

\begin{coqtt}
Inductive \(P\sb{h}\)\(\colon\) \MAT\arr\DIGinf\arr\Propcoq :=
|\(P\sb{hab}\)\(\colon\)\(\forall\)(\(\mu\colon\MAT\))(\(\alpha\colon\DIGinf\)),(\(\forall\)(\(n\colon\NN\)), \(\Psi\sb{h}\) \((n{+}1)\) \(\mu\) \(\alpha\)) \arr\(P\sb{h}\) \(\mu\) \(\alpha\).
\end{coqtt}
The sole constructor of this type ensures that after each emission, which occurs
because of $E_h$, the new \mobius map passed to the homographic algorithm
results in a new emission. This fact is implicit in the following two properties
of $P_h$ that are needed in the definition of the homographic algorithm. First
lemma states the relation between $P_h$ and $E_h$:

\begin{coqtt}
Lemma \(P\sb{h}_E\sb{h}\)\(\colon\)\(\forall\)(\(\mu\colon\MAT\))(\(\alpha\colon\DIGinf\)), \(P\sb{h}\) \(\mu\) \(\alpha\) \arr\(E\sb{h}\) \(\mu\) \(\alpha\).
\end{coqtt}
The second lemma relates $\overline{m}_{h}$ and $P_h$, and shows that $P_h$ is
indeed passed to the future arguments.

\begin{coqtt}
Lemma \(\overline{m}\sb{h}_P\sb{h}\)\(\colon\)\(\forall\)(\(\mu\colon\MAT\))(\(\alpha\colon\DIGinf\))(\(t\colon\!E\sb{h} \mu \alpha\)), 
  let \(\mu'\):=\(\pi\sb{23}\)\((\overline{m}\sb{h} \mu \alpha t)\) in let \(\alpha'\):=\(\pi\sb{33}\)\((\overline{m}\sb{h} \mu \alpha t)\) in 
    \(P\sb{h}\) \(\mu\) \(\alpha\) \arr\(P\sb{h}\) \(\mu'\) \(\alpha'\).
\end{coqtt}

The proof of both of the above lemmas is based on the inverse of the
constructors of \(\Psi\sb{h}\), namely the following lemma which in turn is a
consequence of the Lemma~\ref{lemma:modulus_h_PI}.

\begin{lem}  \label{lemma:Psi_h_inv}
  For all $n$ let $\mu\in\MAT$ $\alpha\in\DIGinf$. 
  \begin{enumerate}[\em(1)]
  \item If $\Psi\sb{h}(n,\mu,\alpha)$ holds then $E\sb{h}(\mu,\alpha)$ holds. \label{item:Psi_h_inv_1}
  \item Let $t$ be a proof that $E\sb{h}(\mu,\alpha)$ holds. Then if $\Psi\sb{h}(n+1,\mu,\alpha)$ holds then 
    $$%\begin{displaymath}
      \Psi\sb{h}\big(n,\pi\sb{23}(\overline{m}\sb{h}(\mu,\alpha,t)),\pi\sb{33}(\overline{m}\sb{h}(\mu,\alpha,t))\big)\enspace.\eqno{\qEd}
    $$%\end{displaymath}
    \label{item:Psi_h_inv_2}
  \end{enumerate}
\end{lem}

Finally we are ready to define the homographic algorithm as a function
\begin{displaymath}
  \map{\bar{h}}{\Pi (\mu\colon \MAT)(\alpha\colon\DIGinf).P_h(\mu,\alpha)}{\DIGinf}
\end{displaymath}
that accommodates the proof of its own productivity as one of its arguments.
Here the \coq keyword \verb|CoFixpoint| denotes that we are using the \cofix
rule (see Section~\ref{sec:coind-type}).

\begin{coqtt}
CoFixpoint \(\bar{h}\) (\(\mu\colon\MAT\)) (\(\alpha\colon\DIGinf\)) (\(p\colon\!P\sb{h} \mu \alpha\)) \(\colon\)\DIGinf :=
  \textcons \(\pi\sb{13}(\overline{m}\sb{h} \mu \alpha (P\sb{h}_E\sb{h} \mu \alpha p))\)
       (\(\bar{h}\) \(\pi\sb{23}(\overline{m}\sb{h} \mu \alpha (P\sb{h}_E\sb{h} \mu \alpha p))\)
          \(\pi\sb{33}(\overline{m}\sb{h} \mu \alpha (P\sb{h}_E\sb{h} \mu \alpha p))\)
          \((\overline{m}\sb{h}_P\sb{h} \mu \alpha (P\sb{h}_E\sb{h} \mu \alpha p) p)\)).
\end{coqtt}
               
This definition passes the guardedness condition of \coq. Thus we have tackled
the problem of productivity by changing the function domain and adding a proof
obligation.

%%%%%%%%%%%%%%%%%%%%%%%%%%%%%%%%%%%%%%%%%%%%%%%%%%%%%%%%%%%%%%%%%%%%%%%%%%%%%%%%
%                                                                              %
%                        Subsection: Cofixed point equations                   %
%                                                                              %
%%%%%%%%%%%%%%%%%%%%%%%%%%%%%%%%%%%%%%%%%%%%%%%%%%%%%%%%%%%%%%%%%%%%%%%%%%%%%%%%
\subsection{Cofixed Point Equations}\label{subsec:cofixed-point}
Next we show that $\bar{h}$ satisfies the specification of the homographic
algorithm. At this point we need to use the extensional equality \bisim on
streams to prove an extensional proof irrelevance for $\bar{h}$. The proof of
this lemma uses Lemma~\ref{lemma:modulus_h_PI}.

\begin{lem} \label{lemma:homographic_EPI} Let $\mu\in\MAT,\alpha\in\DIGinf$ and
  $p, p'$ be two proofs that $P\sb{h}(\mu,\alpha)$ holds. Then the observable
  outcome of $\bar{h}$ is independent of $p$ and $p'$, \ie
  $$%\begin{displaymath}
    \bar{h}(\mu,\alpha,p) \;\bisim\; \bar{h}(\mu,\alpha,p')\enspace.\eqno{\qEd}
  $$%\end{displaymath}\qed
\end{lem}

Subsequently, we use the above lemma together with
Lemma~\ref{lemma:modulus_h_fixedpoints} to prove that $\bar{h}$ satisfies the
specification of the homographic algorithm. We call these the \emph{cofixed
  point equations} of the homographic algorithm because they can be considered
as the dual of the fixed point equations for recursive functions.

\begin{lem} \label{lemma:homographic_cofixedpoints}
  Let $\mu\in\MAT,\alpha\in\DIGinf$ and $p$ be a proof that $P\sb{h}(\mu,\alpha)$
  holds.
  \begin{enumerate}[\em(1)]
  \item If \,\emitting{\mu}{\LL} holds then 
    \begin{displaymath}
      \bar{h}(\mu,\alpha,p) \;\bisim\; \textcons\appl \LL\appl \bar{h}(\LL\sp{-1}\after\mu,\alpha)\enspace.
    \end{displaymath} \label{item:homographic_cofixedpoints_1}
  \item If \,\(\neg\emitting{\mu}{\LL}\) but \emitting{\mu}{\RR} holds then
    \begin{displaymath}
      \bar{h}(\mu,\alpha,p) \;\bisim\; \textcons\appl \RR\appl \bar{h}(\RR\sp{-1}\after\mu,\alpha)\enspace.
    \end{displaymath} \label{item:homographic_cofixedpoints_2}
  \item If \,\(\neg\emitting{\mu}{\LL}\) and \(\neg\emitting{\mu}{\RR}\) but
    \emitting{\mu}{\MM} holds then
    \begin{displaymath}
      \bar{h}(\mu,\alpha,p) \;\bisim\; \textcons\appl \MM\appl \bar{h}(\MM\sp{-1}\after\mu,\alpha)\enspace.
    \end{displaymath} \label{item:homographic_cofixedpoints_3}
  \item If \,\(\neg\emitting{\mu}{\LL}\), \(\neg\emitting{\mu}{\RR}\) and
    \(\neg\emitting{\mu}{\MM}\) holds then for all $p'$ a proof of property
    $P\sb{h}\big(\mu\after(\hd(\alpha)),\tl(\alpha)\big)$ we have
    $$%\begin{displaymath}
      \bar{h}(\mu,\alpha,p) \;\bisim\; \bar{h}\big(\mu\after(\hd(\alpha)),\tl(\alpha),p'\big)\enspace.\eqno{\qEd}
    $$%\end{displaymath} 
    \label{item:homographic_cofixedpoints_4}
    \end{enumerate}
\end{lem}
              
Hence we have shown that our function $\bar{h}$ satisfies the specification of
Section~\ref{sec:algorithms} and is indeed a formalisation of the homographic
algorithm.

So far we have only tackled the formalisation of the homographic algorithm as a
productive coinductive map, and \emph{not} its correctness.  As we stated
earlier the above algorithm (without enforcing any condition on $\mu$) is not
always productive for non-refining \mobius maps. It is important to have in mind
that we have separated the issue of productivity and correctness.  This is in
accordance with separation of \emph{termination} and correctness in the method
of Bove--Capretta for general recursion~\cite{bove.01} or already in the Hoare
logic. Moreover, this separation is also evident in the domain theoretic
semantics of the real numbers~\cite{marcial.07}.

In order to prove the correctness we need to define a suitable semantics (for
example use another model of real numbers) and prove that the effect of the
above algorithm, when applied with a refining \mobius map, is equivalent to the
effect of that \mobius map in \Ibase.  This will be done in Sections
\ref{sec:coind-correctness}--\ref{sec:topologic-correctness}.

%%%%%%%%%%%%%%%%%%%%%%%%%%%%%%%%%%%%%%%%%%%%%%%%%%%%%%%%%%%%%%%%%%%%%%%%%%%%%%%%
%                                                                              %
%                          Quadratic General Corecursive                       %
%                                                                              %
%%%%%%%%%%%%%%%%%%%%%%%%%%%%%%%%%%%%%%%%%%%%%%%%%%%%%%%%%%%%%%%%%%%%%%%%%%%%%%%%
\subsection{Quadratic Algorithm}
In the case of the quadratic algorithm we follow the same method that we used
for the homographic algorithm. We start by defining the inductive type for the
domain of the modulus function.

\begin{coqtt}
Inductive \(E\sb{q}\) \(\colon\) \TEN \arr \DIGinf \arr \DIGinf \arr \Propcoq :=
|\(E\sb{qL}\)\(\colon\)\(\forall\)(\(\xi\colon\TEN\))(\(\alpha \beta\colon\DIGinf\)), \emitting{\xi}{\LL} \arr \(E\sb{q} \xi \alpha \beta\)
|\(E\sb{qR}\)\(\colon\)\(\forall\)(\(\xi\colon\TEN\))(\(\alpha \beta\colon\DIGinf\)), \emitting{\xi}{\RR} \arr \(E\sb{q} \xi \alpha \beta\)
|\(E\sb{qM}\)\(\colon\)\(\forall\)(\(\xi\colon\TEN\))(\(\alpha \beta\colon\DIGinf\)), \emitting{\xi}{\MM} \arr \(E\sb{q} \xi \alpha \beta\)
|\(E\sb{qab}\)\(\colon\)\(\forall\)(\(\xi\colon\TEN\))(\(\alpha \beta\colon\DIGinf\)),
           \(E\sb{q}\) \((\xi\afterL(\hd \alpha))\afterR(\hd \beta)\) (\(\tl \alpha\)) (\(\tl \beta\)) \arr \(E\sb{q} \xi \alpha \beta\).
\end{coqtt}

Using this we define the modulus function by structural recursion on a term of
the above type. Note that in this case the modulus function
\(\overline{m}\sb{q}\) returns a quadruple
$\pair{\phi}{\pair{\xi'}{\pair{\alpha'}{\beta'}}}$ consisting of the emitted
digit, the new quadratic map passed to the continuation of the quadratic
algorithm and the remainder (unabsorbed part) of two the streams of digits.

\begin{coqtt}                                
Fixpoint \(\overline{m}\sb{q}\) (\(\xi\colon\TEN\)) (\(\alpha \beta\colon\DIGinf\)) (\(t\colon\!E\sb{q} \xi \alpha \beta\)) \{struct \(t\)\}
        \(\colon\) \DIG\!\!\!*(\TEN\!\!\!*(\DIGinf\!\!\!*\DIGinf)) :=
match \emittingdec{\xi}{\LL} with
| left _ \(\Rightarrow\) \pair{\LL}{\pair{\LL\sp{-1}\afterM\xi}{\pair{\alpha}{\beta}}}
| right \(t\sb{l}\) \(\Rightarrow\)
   match \emittingdec{\xi}{\RR} with
   | left _ \(\Rightarrow\) \pair{\LL}{\pair{\RR\sp{-1}\afterM\xi}{\pair{\alpha}{\beta}}}
   | right \(t\sb{r}\) \(\Rightarrow\)
      match \emittingdec{\xi}{\MM} with
      | left _ \(\Rightarrow\) \pair{\LL}{\pair{\MM\sp{-1}\afterM\xi}{\pair{\alpha}{\beta}}}
      | right \(t\sb{m}\) \(\Rightarrow\) \(\overline{m}\sb{q}\) \((\xi\afterL(\hd \alpha))\afterR(\hd \beta)\) (\(\tl \alpha\)) (\(\tl \beta\))
                        (\(E\sb{qab}_\textrm{inv}\) \(\xi \alpha \beta\) \(t\sb{l}\) \(t\sb{r}\) \(t\sb{m}\) \(t\))
      end
  end
end.
\end{coqtt}
%\newpage 

Here \(E\sb{qab}\_\textrm{inv}\) is an inverse of the last constructor of the
inductive type $E\sb{q}$ akin to \(E\sb{hab}\_\textrm{inv}\) for the homographic
algorithm. Furthermore we have to prove the proof irrelevance and the fixed
point equations for \(\overline{m}\sb{q}\). For brevity we do not mention them
here but their statement and proofs can be found in~\cite{milad.07}.

Next we define the inductive predicate \(\Psi\sb{q}\) that ensures the validity
of $E\sb{q}$ for finite iterations of $\overline{m}\sb{q}$:

\begin{coqtt}
Inductive \(\Psi\sb{q}\) \(\colon\) \NN \arr \TEN \arr \DIGinf \arr \DIGinf \arr \Propcoq :=
|\(\Psi\sb{q0}\)\(\colon\)\(\forall\)(\(\xi\colon\TEN\)) (\(\alpha \beta\colon\DIGinf\)), \(E\sb{q} \xi \alpha \beta\) \arr \(\Psi\sb{q} 0 \xi \alpha \beta\)
|\(\Psi\sb{qS}\)\(\colon\)\(\forall\)(\(n\colon\NN\))(\(\xi\colon\TEN\)) (\(\alpha \beta\colon\DIGinf\)) (\(t\colon E\sb{q} \xi \alpha \beta\)), 
   \(\Psi\sb{q}\) \(n\) (\(\pi\sb{24}(\overline{m}\sb{q} \xi \alpha \beta t)\)) (\(\pi\sb{34}(\overline{m}\sb{q} \xi \alpha \beta t)\)) (\(\pi\sb{44}(\overline{m}\sb{q} \xi \alpha \beta t)\))\({\arr}\)
                                              \(\Psi\sb{q} (n{+}1) \xi \alpha \beta\).
\end{coqtt}

This allows us to define the productivity predicate $P\sb{q}$:

\begin{coqtt}
Inductive \(P\sb{q}\) \(\colon\) \TEN \arr \DIGinf \arr \DIGinf \arr\Propcoq :=
|\(P\sb{qab}\)\(\colon\)\(\forall\)(\(\xi\colon\TEN\)) (\(\alpha \beta\colon\DIGinf\)), (\(\forall\) (\(n\colon\NN\)), \(\Psi\sb{q} n \xi \alpha \beta\)) \arr \(P\sb{q} \xi \alpha \beta\).
\end{coqtt}

Once again we need to prove two lemmas relating $P_{q}$ with $E_{q}$ and
$\overline{m}\sb{q}$.

\begin{coqtt}
Lemma \(P\sb{q}_E\sb{q}\)\(\colon\)\(\forall\) (\(\xi\colon\TEN\)) (\(\alpha \beta\colon\DIGinf\)),\(P\sb{q} \xi \alpha \beta\) \arr \(E\sb{q} \xi \alpha \beta\).

Lemma \(\overline{m}\sb{q}_P\sb{q}\)\(\colon\) \(\forall\) (\(\xi\colon\TEN\)) (\(\alpha \beta\colon\DIGinf\)) (\(t\colon E\sb{q} \xi \alpha \beta\)), 
   let \(\xi'\):=\(\pi\sb{24}(\overline{m}\sb{q} \xi \alpha \beta t)\) in let \(\alpha'\):=\(\pi\sb{34}(\overline{m}\sb{q} \xi \alpha \beta t)\) in
     let \(\beta'\):=\(\pi\sb{44}(\overline{m}\sb{q} \xi \alpha \beta t)\) in 
       \(P\sb{q} \xi \alpha \beta\) \arr \(P\sb{q} \xi' \alpha' \beta'\).
\end{coqtt}

Finally we can define the quadratic algorithm as a function into the coinductive
type of streams
\begin{displaymath}
 \map{\bar{q}}{\Pi (\xi\colon \TEN)(\alpha\;\beta\colon\DIGinf).
  P_q(\xi,\alpha,\beta)}{\DIGinf} 
\end{displaymath}
using the cofixed point operator of \coq:

\begin{coqtt}
CoFixpoint \(\bar{q}\) (\(\xi\colon\TEN\)) (\(\alpha \beta\colon\DIGinf\)) (\(p\colon\!P\sb{q} \xi \alpha \beta\)) \(\colon\)\DIGinf :=
   \textcons \(\pi\sb{14}(\overline{m}\sb{q} \xi \alpha \beta (P\sb{q}_E\sb{q} \xi \alpha \beta p))\)
        (\(\bar{q}\) \(\pi\sb{24}(\overline{m}\sb{q} \xi \alpha \beta (P\sb{q}_E\sb{q} \xi \alpha \beta p))\)
           \(\pi\sb{34}(\overline{m}\sb{q} \xi \alpha \beta (P\sb{q}_E\sb{q} \xi \alpha \beta p))\) 
           \(\pi\sb{44}(\overline{m}\sb{q} \xi \alpha \beta (P\sb{q}_E\sb{q} \xi \alpha \beta p))\)
           \((\overline{m}\sb{q}_P\sb{q} \xi \alpha \beta (P\sb{q}_E\sb{q} \xi \alpha \beta p) p)\)).
\end{coqtt}

To prove that \(\bar{q}\) satisfies the specification of the quadratic algorithm
we first need the extensional proof irrelevance:

\begin{lem} \label{lemma:quadratic_EPI} Let $\xi\in\TEN,\alpha,\beta\in\DIGinf$
  and $p,p'$ be two proofs that $P\sb{q}(\xi,\alpha,\beta)$ holds. Then the
  observable outcome of $\bar{q}$ is independent of $p$ and $p'$, \ie
  $$%\begin{displaymath}
    \bar{q}(\xi,\alpha,\beta,p) \;\bisim\;
    \bar{q}(\xi,\alpha,\beta,p')\enspace.\eqno{\qEd} 
  $$%\end{displaymath}\qed
\end{lem}

Applying this lemma and the fixed point equations of $\overline{m}\sb{q}$ we can
prove the cofixed point equations of $\bar{q}$.

\begin{lem} \label{lemma:quadratic_cofixedpoints}
  Let $\xi\in\TEN,\alpha,\beta\in\DIGinf$ and $p$ be a proof that
  $P\sb{q}(\xi,\alpha,\beta)$ holds.
  \begin{enumerate}[\em(1)]
  \item If \,\emitting{\xi}{\LL} holds then 
    \begin{displaymath}
      \bar{q}(\xi,\alpha,\beta,p) \;\bisim\; \textcons\appl \LL\appl \bar{q}(\LL\sp{-1}\after\xi,\alpha,\beta)\enspace.
    \end{displaymath} \label{item:quadratic_cofixedpoints_1}
  \item If \,\(\neg\emitting{\xi}{\LL}\) but \emitting{\xi}{\RR} holds then
    \begin{displaymath}
      \bar{q}(\xi,\alpha,\beta,p) \;\bisim\; \textcons\appl \RR\appl \bar{q}(\RR\sp{-1}\after\xi,\alpha,\beta)\enspace.
    \end{displaymath} \label{item:quadratic_cofixedpoints_2}
  \item If \,\(\neg\emitting{\xi}{\LL}\) and \(\neg\emitting{\xi}{\RR}\) but
    \emitting{\xi}{\MM} holds then
    \begin{displaymath}
      \bar{q}(\xi,\alpha,\beta,p) \;\bisim\; \textcons\appl \MM\appl \bar{q}(\MM\sp{-1}\after\xi,\alpha,\beta)\enspace.
    \end{displaymath} \label{item:quadratic_cofixedpoints_3}
  \item If \,\(\neg\emitting{\xi}{\LL}\), \(\neg\emitting{\xi}{\RR}\) and
    \(\neg\emitting{\xi}{\MM}\) holds then for all $p'$ a proof of property
    $P\sb{q}\big((\xi\afterL(\hd(\alpha)))\afterR(\hd(\beta)),\tl(\alpha),\tl(\beta)\big)$
    we have
    $$%\begin{displaymath}
      \bar{q}(\xi,\alpha,\beta,p) \;\bisim\; \bar{q}\big(\xi\afterL(\hd(\alpha)))\afterR(\hd(\beta),\tl(\alpha),\tl(\beta),p'\big)\enspace.\eqno{\qEd}
    $$%\end{displaymath}
    \label{item:quadratic_cofixedpoints_4}
  \end{enumerate}
\end{lem}

Hence $\bar{q}$ agrees with the specification of the quadratic algorithm.

%%%%%%%%%%%%%%%%%%%%%%%%%%%%%%%%%%%%%%%%%%%%%%%%%%%%%%%%%%%%%%%%%%%%%%%%%%%%%%%%
%                                                                              %
%                              Analysis of the method                          %
%                                                                              %
%%%%%%%%%%%%%%%%%%%%%%%%%%%%%%%%%%%%%%%%%%%%%%%%%%%%%%%%%%%%%%%%%%%%%%%%%%%%%%%%
\subsection{General Corecursion?}\label{subsec:general-corec?}
Evidently the method for formalising the quadratic algorithm mimics precisely
the one used for the homographic algorithm. This suggests that one can
generalise this method to obtain a scheme in style of~\cite{cancila.03} for
formalising specification of partial functions on coinductive types.  Such a
method would be the dual of the Bove--Capretta~\cite{bove.01} for general
recursion. For our situation the dual term \emph{general corecursion} seems
suitable.  In this article we have not developed such a scheme, as our focus
lies on the special case of exact arithmetic algorithms for the coinductive type
of reals.  Nevertheless, all the intermediate inductive predicates and recursive
functions can be obtained by following the shape of the specification. Therefore
we consider the method to be generic enough for formalising arbitrary partial
coalgebra maps for strictly positive functors in any category modelling \CIC.

In fact the method might work in categories for simpler extensions of \martinlof
type theory. This is because the method does not rely on properties peculiar to
\CIC; even the distinction between \Setcoq and \Propcoq is not necessary and we
could put all the inductive predicates in \Setcoq. However, with an eye on
program extraction, we prefer to keep the distinction between informative and
non-informative objects. Note that if we extract the function $\bar{h}$ the
argument $P\sb{h}(\mu,\alpha)$ will be discarded, resulting in a function
$\map{\hat{h}}{\MAT\times\DIGinf}{\DIGinf}$ which is only different from the
original specification modulo unfolding (see the discussion by
Bertot~\cite{bertot.05}).

It remains to be seen whether the method can be applied in categories other than
those modelling some extensions of \martinlof type theory.

Comparing our method with the one given by Bertot~\cite{bertot.05} we observe
that both there and in our work the same idea of dualising Bove--Capretta's
method is pursued.  One difference between our work and~\cite{bertot.05} is that
we consider $\Psi_{h}$ to be an inductive type while Bertot uses a coinductive
predicate \textsf{F\_infinite}. But our predicate $P_{h}$ (which is a wrapper
for $\Psi_{h}$) and Bertot's \textsf{F\_infinite} seem to be extensionally
equal.  Moreover we need the inductive predicate $\Psi_{h}$ to capture the
iteration of $\overline{m}\sb{h}$, a characteristic that does not occur in
Bertot's method for \filter. This is due to the slight difference between the
homographic algorithm and the general form of \filter function: in the
homographic algorithm the property \emittingvoid is a dynamic property because
the \mobius map, being passed to future steps of the function, is changing all
the time; therefore the property that states the productivity should keep track
of this.  However, by considering a more dynamic form of \filter, such as the
function \textsf{etree\_filter} introduced in~\cite[p.~128]{milad.04b} it might
be possible to extend the method of~\cite{bertot.05} and apply it in our case.

Another notable difference is our use of bisimulation equality and the proofs
for extensional cofixed point equations which are not present
in~\cite{bertot.05} where instead another coinductive predicate is used to
describe the \emph{connectedness} of a stream with respect to a given property.

Finally, we remark that the \role of the productivity predicate in our work is
reminiscent of the \texttt{forall} function in~\cite{simpson.98} (which is
attributed to Berger).  There, this function is employed to provide a universal
quantifier for total predicates on streams and is used for obtaining higher
order functions such as the numerical integration. However, this function which
is the basis for defining other functions in~\cite{simpson.98}, itself does not
satisfy the \coq guardedness condition and hence its formalisation in \coq will
require additional trickery similar to what we did here for our algorithms.  On
the other hand, in our work the productivity predicates are inductively defined
data-types rather than functions and hence are not hampered by the guardedness
condition.  It might, however, be possible to combine our method with the
techniques in~\cite{simpson.98} for defining higher order functions.

%%%%%%%%%%%%%%%%%%%%%%%%%%%%%%%%%%%%%%%%%%%%%%%%%%%%%%%%%%%%%%%%%%%%%%%%%%%%%%%%
%                                                                              %
%                              Section: Representation                         %
%                                                                              %
%%%%%%%%%%%%%%%%%%%%%%%%%%%%%%%%%%%%%%%%%%%%%%%%%%%%%%%%%%%%%%%%%%%%%%%%%%%%%%%%
\section{Representation}\label{sec:representation}
As it is the case with all algorithms, `to prove the correctness' of the
homographic and quadratic algorithms can point to different concepts:
\begin{itemize}
\item[(i)] To prove that the algorithms satisfy their \haskell-like specification. 
\item[(ii)] To prove that the algorithms turn the set \DIGinf to a partial field
  and behave as \mobius and quadratic maps on this partial field.
\item[(iii)] To prove that the algorithms correspond to \mobius and quadratic
  maps on \Ibase.
\end{itemize}

Concept (i) tantamounts to proving the cofixed point equations and was carried
out in Section~\ref{subsec:cofixed-point}. Concept (ii) requires that we focus
on the field operations (via specific tensors for $+,\times$) and prove that
they satisfy the algebraic properties of field operations such as commutativity
and distributivity.  Concept (iii) requires the use of a model of real numbers
and indicates that we will project the algorithm to functions on this standard
model.  It is clear that (iii) is much less work, as we only have to prove the
correspondence of the algorithms once and can reduce every question on \DIGinf
to a question on the standard model of \RRR.  This way we do not have to prove
one-by-one the field axioms for \DIGinf. The remainder of this work is based on
the concept (iii).

To prove that the algorithms are correct in the sense of (iii), first we should
prove that every stream in \DIGinf represents a real number in \Ibase. This
means that there exists a total\footnote{In fact \DIGinf is a
  \emph{representation} which means $\rho$ is also surjective.  This is easily
  provable~\cite[\S~5]{milad.04b} but it is not needed in the correctness proofs
  for our algorithms.}  map $\rho$ from $\DIGinf$ to \Ibase such that for all
$\phi_0\phi_1\dots\in\DIGinf$ we have
\begin{displaymath}
  \set{\rho(\phi_0\phi_1\dots)}=\bigcap\limits_{i=1}^{\infty}\phi_0\after\ldots \phi_{i}(\Ibase)\enspace. 
\end{displaymath}
This can be proven by coinduction, but one needs to define a coinductive
predicate that captures the existence of $\rho$. This leads to the following
definition for a binary predicate \map{\rep}{\DIGinf\times\Ibase}{\Propcoq} with the intended meaning that 
$\rep(\alpha,r)$ holds if $\rho(\alpha)=\{r\}$.

\begin{coqtt}
CoInductive \rep \(\colon\) \DIGinf \arr \RRR \arr \Propcoq :=
  | \(\rep\sb{L}\) \(\colon\) \(\forall\) (\(\alpha \beta\colon\DIGinf\)) (\(r\colon\RRR\)), \(-1{\leq}r{\leq}1\)\arr 
                \(\rep \alpha r\) \arr \(\beta\) \bisim \textcons \(\LL \alpha\) \arr \(\rep \beta (r-1)/(r+3)\)
  | \(\rep\sb{R}\) \(\colon\) \(\forall\) (\(\alpha \beta\colon\DIGinf\)) (\(r\colon\RRR\)), \(-1{\leq}r{\leq}1\)\arr 
               \(\rep \alpha r\) \arr \(\beta\) \bisim \textcons \(\RR \alpha\) \arr \(\rep \beta (r+1)/(-r+3)\)
  | \(\rep\sb{M}\) \(\colon\) \(\forall\) (\(\alpha \beta\colon\DIGinf\)) (\(r\colon\RRR\)), \(-1{\leq}r{\leq}1\)\arr 
                         \(\rep \alpha r\) \arr \(\beta\) \bisim \textcons \(\MM \alpha\) \arr \(\rep \beta r/3\).
\end{coqtt}
The constructors of this coinductive predicate spell out the effect of each
digit and as such depend on the choice of the digits. However, they can easily
be adapted or generalised for working with other digit sets. The predicate is
similar to the predicate \texttt{represents} of
Bertot~\cite{bertot.05a,bertot.06} and (to a lesser extent) to the predicate
$\sim'$ of Hou~\cite{hou.06} but has a notable difference: the clause
$\beta\bisim \textcons\;\;d\;\;\alpha$ that is added to each constructor.  The
purpose of this clause is to facilitate the use of cofixed point equations.
Without this clause \rep would still have the intended topological semantics in
terms of $\rho$, but it would not be usable in the coinductive proof of
correctness that we intend to give in the next section. The reason is due to the
guardedness condition of \coq: even without the \bisim clause in the
constructors of \rep we could find a proof $X$, by coinduction, for the property
that
\begin{equation}\label{eq:rep_1}
  \forall \alpha\beta r,\;\;\rep(\alpha,r) \rightarrow \alpha\bisim\beta \rightarrow \rep(\beta,r)\enspace.
\end{equation} 
This is the basic property of \rep that \emph{should have been} enough for the
correctness proof. But upon rewriting (\ref{eq:rep_1}) in the course of
coinductive proof $\Delta$ of correctness we would violate the guardedness
condition.  This would happen because we would have supplied a recursive
occurrence of the coinductive proof $\Delta$ which occurs in a subterm of the
form
\begin{displaymath}
  X\appl \alpha_0\appl \beta_0\appl r_0 \appl (\rep{\sb{\phi}}\appl \Delta)
\end{displaymath}
(where $\rep{\sb{\phi}}$ is a constructor of \rep).  In such a situation $\Delta$
is guarded by $\rep{\sb{\phi}}$ and $X$. This does not satisfy the guardedness
condition because $X$ is itself a cofixed point whose expansion takes the
coinductive proof $\Delta$ as an argument in its recursive occurrence in a way
that the guardedness condition is rejected. Using cofixed point equations
instead of (\ref{eq:rep_1}) we will not land in this situation.  Thus we have
decided to add the \bisim clause which will eliminate the need for
(\ref{eq:rep_1}) and instead use the cofixed point equations in the correctness
proofs.

Note that (\ref{eq:rep_1}) is still a correct statement and can be used in other
situations. In fact we can use it to prove that the inverse of constructors of
\rep hold, \eg:
\begin{equation}\label{eq:rep_10}
  \forall \alpha r,\;\;\rep(\textcons\;\;\LL\;\;\alpha,r) \rightarrow \rep(\alpha,\frac{3r+1}{-r+1})\enspace.
\end{equation}

The inversion lemmas in turn are used in proving the link between a stream and
its future tails. Let $\alpha_n$ (\resp $\alpha|_{n}$) denote the $n+1$-st digit
of $\alpha$ (\resp the stream obtained by dropping the first $n$ digits of
$\alpha$).  Then we can prove by induction on $n$ and using the inversion lemmas
that
\begin{equation}\label{eq:rep_2}
  \forall \alpha r,\;\;\rep(\alpha,r) \rightarrow \rep(\alpha|_{n},\alpha_{n-1}^{-1}\after\ldots \alpha_{0}^{-1}(r))\enspace.
\end{equation}

To show that \rep satisfies its metric property we have to define a function
\decode{\_} that evaluates a stream and obtains the real number which is
represented by it (\cf \texttt{real\_value} in~\cite{bertot.06}). In fact this
function calculates the limit of converging sequence of shrinking intervals that
is obtained by successive application of the digits starting from the base
interval. To be able to define \decode{\_} we should show this converging
property.  This proof is directly dependent on the metric properties of the
specific digit set that we have chosen. Setting
$\operatorname{\mathsf{diam}}([a,b])=b-a$ we have to show that
\begin{equation} \label{eq:rep_0}
  \max\set{\operatorname{\mathsf{diam}}\big({\phi_0\after \phi_1\after\ldots \phi_{k-1}(\Ibase)\big)}|\phi_i\in\DIG}\leq \frac{2}{k+1}\enspace.
\end{equation}
This is provable by induction on $k$~\cite[Corollary~5.7.9]{milad.04b} and it
entails that the diameters of the intervals form a Cauchy sequence, and so do
their endpoints. Hence if we define $l_k(\alpha)$ (\resp $u_k(\alpha)$) to be
the lower bound (\resp upper bound) of the interval $\alpha_0\after
\alpha_1\after\ldots \alpha_{k-1}(\Ibase)$ we can define\footnote{Note that we
  could have equivalently used the upper bounds.}
\begin{displaymath}
\decode{\alpha}=\lim\limits_{i\arr\infty}l_i(\alpha)\enspace.
\end{displaymath}
Note that (\ref{eq:rep_0}) can be rewritten as
\begin{equation}\label{eq:rep_3}
  \forall \alpha k, u_k(\alpha)-l_k(\alpha)\leq \frac{2}{k+1}\enspace;
\end{equation}
and we can prove (by induction on $k$) that
\begin{equation}\label{eq:rep_11}
  \forall \alpha k r, \rep(\alpha,r) \rightarrow r\in[l_k(\alpha),u_k(\alpha)]\enspace;
\end{equation}
and hence
\begin{equation}\label{eq:rep_8}
 \forall \alpha k r, \rep(\alpha,r) \rightarrow r\in\Ibase\enspace.
\end{equation}
Furthermore using the properties of limit we can prove for $\phi$ a digit
\begin{align}
&\decode{\textcons\;\; \phi\;\; \alpha}=\phi(\decode{\alpha})\enspace,\label{eq:rep_4}\\
&\decode{\alpha}\in\Ibase\enspace.\label{eq:rep_5}
\end{align}
Thus we can prove the following by an easy coinduction on the structure of \rep.
\begin{equation}\label{eq:rep_9}
\forall\alpha, \rep(\alpha,\decode{\alpha})\enspace.
\end{equation}
Finally we can prove the main properties of \rep
\begin{align}
& \forall \alpha r,\;  \decode{\alpha}=r \rightarrow \rep(\alpha,r)\enspace;\label{eq:rep_7}\\
&\forall \alpha r,\;  \rep(\alpha,r) \rightarrow \decode{\alpha}=r\enspace.\label{eq:rep_6}
\end{align}
The proof of~(\ref{eq:rep_7}) follows from (\ref{eq:rep_9}) and
(\ref{eq:rep_5}) while (\ref{eq:rep_6}) needs in addition some properties of the
limit.  

Hence we have shown that \rep satisfies its intended metric property with
respect to the map $\rho$ defined in the beginning of this section.  We conclude
the section by pointing out what \rep does \emph{not} entail. The most important
aspect is that our representation \DIGinf is an \emph{admissible}
representation, \ie it contains enough redundancy so that the usual computable
functions are computable with respect to this
representation~\cite[Corollary~5.7.10]{milad.04b}. However, the \bisim equality
does not know anything about this redundancy and it distinguishes the two
streams representing the same real number. Therefore for two different
representations $\alpha_1,\alpha_2$ of a real number $r$, there are two
different proofs $\rep(\alpha_1,r)$ and $\rep(\alpha_2,r)$ that do not have any
syntactic relation with each other. This, of course, is not an issue for our
application of \rep in the correctness proofs of the next section.

%%%%%%%%%%%%%%%%%%%%%%%%%%%%%%%%%%%%%%%%%%%%%%%%%%%%%%%%%%%%%%%%%%%%%%%%%%%%%%%%
%                                                                              %
%                            Coinductive Correctness                           %
%                                                                              %
%%%%%%%%%%%%%%%%%%%%%%%%%%%%%%%%%%%%%%%%%%%%%%%%%%%%%%%%%%%%%%%%%%%%%%%%%%%%%%%%
\section{Coinductive Correctness}\label{sec:coind-correctness}

We are going to prove that the homographic and quadratic algorithms correspond
to \mobius and quadratic maps on \Ibase as a subset of the standard model of
\RRR.  We base our correctness proofs on the coinductive predicate \rep and we
prove that for the functions $\bar{h}$ and $\bar{q}$ of
Section~\ref{sec:general-corec} we have
\begin{align}
  &\forall \mu\alpha p r,\;\; \rep(\alpha,r) \rightarrow \rep (\bar{h}(\mu,\alpha,p), \mu(r))\enspace;\label{eq:correct_1}\\
  &\forall \xi\alpha\beta p r_1 r_2,\;\; \rep(\alpha,r_1) \rightarrow
  \rep(\alpha,r_2) \rightarrow \rep (\bar{q}(\xi,\alpha,\beta,p),
  \xi(r_1,r_2))\enspace.\label{eq:correct_2}
\end{align}
It is clear that once we have proven these, applying the
Properties~(\ref{eq:rep_7})--(\ref{eq:rep_6}) of \rep, we can derive
\begin{displaymath}
\begin{array}{c}
  \forall \mu\alpha p r,\;\; \decode{\alpha}=r \rightarrow \decode{\bar{h}(\mu,\alpha,p)}=\mu(r)\enspace;\\
  \forall \xi\alpha\beta p r_1 r_2,\;\; \decode{\alpha}=r_1 \rightarrow \decode{\beta}=r_2
  \rightarrow \decode{\bar{q}(\xi,\alpha,\beta,p)}=\xi(r_1,r_2)\enspace.
\end{array}
\end{displaymath}

Note that these statements require a proof obligation of the productivity
predicates $P_h$ and $P_q$, following our definition of $\bar{h}$ and $\bar{q}$.
This means that we prove the correctness modulo the existence of proofs of these
predicates. In the remainder of this section we show how to prove
(\ref{eq:correct_1}) and (\ref{eq:correct_2}).

%%%%%%%%%%%%%%%%%%%%%%%%%%%%%%%%%%%%%%%%%%%%%%%%%%%%%%%%%%%%%%%%%%%%%%%%%%%%%%%%
%                                                                              %
%                      Homographic Coinductive Correctness                     %
%                                                                              %
%%%%%%%%%%%%%%%%%%%%%%%%%%%%%%%%%%%%%%%%%%%%%%%%%%%%%%%%%%%%%%%%%%%%%%%%%%%%%%%%
\subsection{Homographic Algorithm}\label{subsec:homographic-coind-correctness}
We want to prove (\ref{eq:correct_1}). This means that in addition to
$\mu,\alpha$ and $r$ we are also given a proof $p$ of the statement
$P\sb{h}(\mu,\alpha)$ that ensures the productivity of $\bar{h}(\mu)$ at
$\alpha$. We use $p$ to obtain some auxiliary tools that we will need in the
proof of (\ref{eq:correct_1}).  We will also use the terms that were used in our
technique for general corecursion (see Section~\ref{sec:general-corec}).  First
we need a function
\begin{displaymath}
 \map{\overline{\delta}_h}{\Pi(\mu\colon\MAT)(\alpha\colon\DIGinf).  E\sb{h}(\mu,\alpha)}{\NN}
\end{displaymath}
that counts the number of absorption steps before the first (eventually) coming
emission step. Note the resemblance with the definition of the modulus of
productivity \(\overline{m}\sb{h}\).

\begin{coqtt}
Fixpoint \(\overline{\delta}\sb{h}\) (\(\mu\colon\MAT\))(\(\alpha\colon\DIGinf\))(\(t\colon\!E\sb{h} \mu \alpha\))\{struct \(t\)\}\(\colon\)\NN:=
match \emittingdec{\mu}{\LL} with
| left _\(\Rightarrow\) 0
| right \(t\sb{l}\)\(\Rightarrow\)
  match \emittingdec{\mu}{\RR} with
  | left _\(\Rightarrow\) 0
  | right \(t\sb{r}\)\(\Rightarrow\)
    match \emittingdec{\mu}{\MM} with
    | left _\(\Rightarrow\) 0
    | right \(t\sb{m}\)\(\Rightarrow\)\(1{+}\overline{\delta}\sb{h}\) (\(\mu\after(\hd \alpha)\)) (\(\tl\) \(\alpha\)) (\(E\sb{hab}\_\textrm{inv}\) \(\mu\) \(\alpha\) \(t\sb{l}\) \(t\sb{r}\) \(t\sb{m}\) \(t\))
    end
  end
end.
\end{coqtt}
We will also need to prove the proof irrelevance of \(\overline{\delta}\sb{h}\)
(\ie its value is independent of $t$), its fixed point equation and its
relationship with \(\overline{m}\sb{h}\).  We state the latter:

\begin{lem} \label{lemma:delta_h_modulus_h} Let $\mu\in\MAT, \alpha\in\DIGinf$ and
  $t$ be a proof that $E\sb{h}(\mu,\alpha)$ holds. Then for all $n$ if
  \(\overline{\delta}\sb{h}(\mu,\alpha,t)=n\) then there exists $\phi\in\DIG$
  such that
$$%\begin{displaymath}
\overline{m}\sb{h}(\mu,\alpha,t)=\pair{\phi}{\pair{\phi\sp{-1}{\after}\mu{\after}\alpha\sb{0}{\after}\ldots{\after}\alpha\sb{n-1}}{\alpha|\sb{n}}}\enspace.\eqno{qEd}
$$%\end{displaymath}\qed
\end{lem}

Then we need to prove that if the value of $\overline{\delta}$ is $n$ then after
$n$ steps emission will occur, \ie the emission condition will be satisfied:

\begin{lem} \label{lemma:delta_h_emttingvoid} Let $\mu\in\MAT, \alpha\in\DIGinf$
  and $t$ be a proof that $E\sb{h}(\mu,\alpha)$ holds. Then for all $n$ if
  \(\overline{\delta}\sb{h}(\mu,\alpha,t)=n\) then one of the following three
  cases always holds.
  \begin{itemize}
  \item[(a)] \(\;\emitting{\mu{\after}\alpha\sb{0}{\after}\ldots{\after}\alpha\sb{n-1}}{\LL} \wedge \pi\sb{13}\big(\overline{m}\sb{h}(\mu,\alpha,t)\big)=\LL\enspace;\)
  \item[(b)] \(\;\emitting{\mu{\after}\alpha\sb{0}{\after}\ldots{\after}\alpha\sb{n-1}}{\RR} \wedge \pi\sb{13}\big(\overline{m}\sb{h}(\mu,\alpha,t)\big)=\RR\enspace;\)
  \item[(c)] \(\;\emitting{\mu{\after}\alpha\sb{0}{\after}\ldots{\after}\alpha\sb{n-1}}{\MM} \wedge \pi\sb{13}\big(\overline{m}\sb{h}(\mu,\alpha,t)\big)=\MM\enspace.\)\qed
  \end{itemize}
\end{lem}
      
Both lemmas above are proven by induction on $n$.  All this machinery is used in
proving the following lemma which describes the observable (hence the use of
\bisim) situation of the homographic algorithm at the moment of emission. It
explicitly mentions the new input \mobius map passed to the homographic
algorithm, the emission condition and the necessary proof obligation.

\begin{lem} \label{lemma:homographic_emitting} Let $\mu\in\MAT, \alpha\in\DIGinf$
  and $p$ be a proof that $P\sb{h}(\mu,\alpha)$ holds. Then there exist
  $n\in\NN$ and $\phi\in\DIG$ that satisfy the following three conditions.
  \begin{enumerate}[\em(1)]
  \item \(P\sb{h}(\phi\sp{-1}{\after}\mu{\after}\alpha\sb{0}{\after}\ldots{\after}\alpha\sb{n-1},\alpha|\sb{n})\enspace;\)
  \item \(\emitting{\mu{\after}\alpha\sb{0}{\after}\ldots{\after}\alpha\sb{n-1}}{\phi}\enspace;\)
  \item If $p'$ is a proof that
    $P\sb{h}(\phi\sp{-1}{\after}\mu{\after}\alpha\sb{0}{\after}\ldots{\after}\alpha\sb{n-1},\alpha|\sb{n})$
    holds then
    $$%\begin{displaymath}
      \overline{h}(\mu,\alpha,p) \;\bisim\; \textcons\appl\phi\appl \overline{h}(\phi\sp{-1}{\after}\mu{\after}\alpha\sb{0}{\after}\ldots{\after}\alpha\sb{n-1},\alpha|\sb{n},p')\enspace.\eqno{\qEd}
    $$%\end{displaymath}\qed
\end{enumerate}
\end{lem}

Finally we need a property of refining \mobius maps whose proof is
immediate, but we state it explicitly to highlight its use.
\begin{lem}\label{lemma:refining-inverse}
  If \emitting{\mu}{\phi} then $\phi^{-1}\after\mu$ is refining.\qed
\end{lem}

Now we have the necessary tools for proving the correctness of the homographic
algorithm:
\begin{thm}\label{theorem:hcorrectness}
  Let $\mu\in\MAT$, $\alpha\in\DIGinf$, $r\in\RRR$ and let $p$ be a proof that
  $P\sb{h}(\mu,\alpha)$ holds. If $\rep(\alpha,r)$ holds then
\begin{displaymath}
\rep(\bar{h}(\mu,\alpha,p), \mu(r))\enspace.
\end{displaymath}
\end{thm}
\proof By Lemma~\ref{lemma:homographic_emitting} there exist $n$, $\phi$ and $p'$
such that
\begin{gather}
\emitting{\mu{\after}\alpha\sb{0}{\after}\ldots{\after}\alpha\sb{n-1}}{\phi}\enspace,\label{eq:hcorrect_1}\\
\overline{h}(\mu,\alpha,p)  \;\bisim\; \textcons\;\; \phi\;\;\; \overline{h}(\phi\sp{-1}{\after}\mu{\after}\alpha\sb{0}{\after}\ldots{\after}\alpha\sb{n-1},\alpha|\sb{n},p')\enspace.\label{eq:hcorrect_2}
\end{gather}

By Property~(\ref{eq:rep_2}) of \rep we have
\begin{equation*}\label{eq:hcorrect_3}
  \rep(\alpha|_{n},\alpha_{n-1}^{-1}\after\ldots \alpha_{0}^{-1}(r))\enspace.
\end{equation*}

Whence by coinduction applied to
\begin{align*}
&\mu_{c}:=\phi\sp{-1}{\after}\mu{\after}\alpha\sb{0}{\after}\ldots{\after}\alpha\sb{n-1}\enspace,\\
&\alpha_{c}:=\alpha|\sb{n} \qquad\qquad p_{c}:=p'\enspace,\\
&r_{c}:=\alpha_{n-1}^{-1}{\after}\ldots \alpha_{0}^{-1}(r)\enspace;
\end{align*}
we obtain $\rep(\bar{h}(\mu_{c},\alpha_{c},p_{c}), \mu_{c}(r_{c}))$, \ie
\begin{equation}\label{eq:hcorrect_4}
\rep\big(\overline{h}(\phi\sp{-1}{\after}\mu{\after}\alpha\sb{0}{\after}\ldots{\after}\alpha\sb{n-1},\alpha|\sb{n},p'),\phi\sp{-1}{\after}\mu{\after}\alpha\sb{0}{\after}\ldots{\after}\alpha\sb{n-1}{\after}\alpha_{n-1}^{-1}{\after}\ldots \alpha_{0}^{-1}(r)\big)\enspace.
\end{equation}

Let $r_1:=\mu_{c}{\after}\alpha_{n-1}^{-1}{\after}\ldots \alpha_{0}^{-1}(r)$.
According to Lemma~\ref{lemma:refining-inverse}, from (\ref{eq:hcorrect_1}) it
follows that $\mu_{c}$ is refining.  Note that by Properties (\ref{eq:rep_8}) and
(\ref{eq:rep_2}) of \rep we have
\begin{displaymath}
\alpha_{n-1}^{-1}{\after}\ldots \alpha_{0}^{-1}(r)\in\Ibase\enspace;
\end{displaymath}
and thus according to the refining property $r_1\in\Ibase$.
                                       
From here and (\ref{eq:hcorrect_4}), according to the statement of the
constructor $\rep_{\phi}$ of \rep applied to $r_1$ and
\begin{align*}
&\overline{h}(\phi\sp{-1}{\after}\mu{\after}\alpha\sb{0}{\after}\ldots{\after}\alpha\sb{n-1},\alpha|\sb{n},p')\enspace,\\
&\overline{h}(\mu,\alpha,p)\enspace;\\
\end{align*}
we obtain 
\begin{equation}\label{eq:hcorrect_5}
\rep\big(\bar{h}(\mu,\alpha,p), \phi(\phi\sp{-1}{\after}\mu{\after}\alpha\sb{0}{\after}\ldots{\after}\alpha\sb{n-1}{\after}\alpha_{n-1}^{-1}{\after}\ldots \alpha_{0}^{-1}(r))\big)\enspace;
\end{equation}
(note that~(\ref{eq:hcorrect_2}) satisfies the \bisim clause in $\rep_{\phi}$).

Finally, by simple rewriting and cancelling out the inverse matrices
in~(\ref{eq:hcorrect_5}) we obtain the conclusion:
\begin{multline*}
  \rep\big(\bar{h}(\mu,\alpha,p), \phi(\phi\sp{-1}{\after}\mu{\after}\alpha\sb{0}{\after}\ldots{\after}\alpha\sb{n-1}{\after}\alpha_{n-1}^{-1}{\after}\ldots \alpha_{0}^{-1}(r))\big) \\
 \hfill =\rep(\bar{h}(\mu,\alpha,p), \phi{\after}\phi\sp{-1}{\after}\mu{\after}\alpha\sb{0}{\after}\ldots{\after}\alpha\sb{n-1}{\after}\alpha_{n-1}^{-1}{\after}\ldots \alpha_{0}^{-1}(r))\\
  \hfill =\rep(\bar{h}(\mu,\alpha,p), \mu(r))\enspace.
\end{multline*}
\qed

%%%%%%%%%%%%%%%%%%%%%%%%%%%%%%%%%%%%%%%%%%%%%%%%%%%%%%%%%%%%%%%%%%%%%%%%%%%%%%%%
%                                                                              %
%                       Quadratic Coinductive Correctness                      %
%                                                                              %
%%%%%%%%%%%%%%%%%%%%%%%%%%%%%%%%%%%%%%%%%%%%%%%%%%%%%%%%%%%%%%%%%%%%%%%%%%%%%%%%
\subsection{Quadratic Algorithm}\label{subsec:quadratic-coind-correctness}
The procedure for the correctness of the quadratic algorithm is quite similar to
the case of the homographic algorithm, only the proof itself is more meticulous.
First we define a function
\map{\overline{\delta}\sb{q}}{\Pi(\xi\colon\TEN)(\alpha,\beta\colon\DIGinf).E\sb{q}(\xi,\alpha,\beta)}{\NN}
that outputs the number of steps to the next emission step. We can prove the
properties similar to those of \(\overline{\delta}\sb{h}\).

The main auxiliary lemma in this case is the following.

\begin{lem} \label{lemma:quadratic_emitting} Let $\xi\in\TEN,
  \alpha,\beta\in\DIGinf$ and $p$ be a proof that $P\sb{q}(\xi,\alpha,\beta)$
  holds. Then there exist $n\in\NN$ and $\phi\in\DIG$ that satisfy the following
  three conditions.
  \begin{enumerate}[\em(1)]
  \item \(P\sb{q}(\phi\sp{-1}{\afterM}\xi\pair{\alpha\sb{0}{\after}\ldots{\after}\alpha\sb{n-1}}{\beta\sb{0}{\after}\ldots{\after}\beta\sb{n-1}},\alpha|\sb{n},\beta|\sb{n})\enspace;\)
  \item \(\emitting{\xi\pair{\alpha\sb{0}{\after}\ldots{\after}\alpha\sb{n-1}}{\beta\sb{0}{\after}\ldots{\after}\beta\sb{n-1}}}{\phi}\enspace;\)
  \item If $p'$ is a proof that
    $P\sb{q}(\phi\sp{-1}{\afterM}\xi\pair{\alpha\sb{0}{\after}\ldots{\after}\alpha\sb{n-1}}{\beta\sb{0}{\after}\ldots{\after}\beta\sb{n-1}},\alpha|\sb{n},\beta|\sb{n})$
    holds, then
    $$%\begin{displaymath}
      \overline{q}(\xi,\alpha,\beta,p) \;\bisim\;
          \textcons\appl \phi\appl \overline{q}(\phi\sp{-1}{\afterM}\xi\pair{\alpha\sb{0}{\after}\ldots{\after}\alpha\sb{n-1}}{\beta\sb{0}{\after}\ldots{\after}\beta\sb{n-1}},\alpha|\sb{n},\beta|\sb{n},p')\enspace.\eqno{\qEd}
    $$%\end{displaymath}\qed
\end{enumerate}
\end{lem}

Note that
$\xi\pair{\alpha\sb{0}{\after}\ldots{\after}\alpha\sb{n-1}}{\beta\sb{0}{\after}\ldots{\after}\beta\sb{n-1}}$
denotes the new tensor after $n$ absorption steps, \ie after $n$ applications of
\afterL and \afterR.

We also need a result on refining tensors which is immediately provable from the
definition of refining and \emittingvoid.
\begin{lem} \label{lemma:quadratic-refining-inverse}
  If \emitting{\xi}{\phi} then $\phi^{-1}\afterM\xi$ is a
  refining tensor.\qed
\end{lem}

From these we can prove the correctness of the quadratic algorithm. In
particular we do \emph{not} need any additional property of \rep apart from
those that were used for the homographic algorithm. The proof is quite similar
to the proof of Theorem~\ref{theorem:hcorrectness} and is formalised in
\coq~\cite{milad.07}, and so we do not detail the proof here.
\begin{thm}\label{theorem:qcorrectness}
  Let $\xi\in\TEN$, $\alpha,\beta\in\DIGinf$, $r_1,r_2\in\RRR$ and let $p$ be a
  proof that $P\sb{q}(\mu,\alpha,\beta)$ holds. If $\rep(\alpha,r_1)$ and
  $\rep(\beta,r_2)$ hold then
$$%\begin{displaymath}
\rep(\bar{q}(\xi,\alpha,\beta,p), \xi(r_1,r_2))\enspace.\eqno{\qEd}
$$%\end{displaymath}
\end{thm}

Note that the above theorems require the existence of proofs for productivity
statements $P\sb{h}$ and $P\sb{q}$.  Deriving this property depends on the
specific metric properties of each tensor and \mobius map. Next we should prove
that for refining maps we can dispose of these productivity predicates.

%%%%%%%%%%%%%%%%%%%%%%%%%%%%%%%%%%%%%%%%%%%%%%%%%%%%%%%%%%%%%%%%%%%%%%%%%%%%%%%%
%                                                                              %
%                            Topological Correctness                           %
%                                                                              %
%%%%%%%%%%%%%%%%%%%%%%%%%%%%%%%%%%%%%%%%%%%%%%%%%%%%%%%%%%%%%%%%%%%%%%%%%%%%%%%%
\section{Final step: Refining, Productivity and Topological Correctness}\label{sec:topologic-correctness}
So far we have shown that the homographic and quadratic algorithms are `correct'
modulo the existence of the productivity predicates $P_h$ and $P_q$. In this
section we will prove that if a \mobius map (\resp quadratic map) is refining
then irrespective of the used input stream(s) the property $P_h$ (\resp $P_q$)
is always satisfied.

being refining in enough to ensure the correctness. In light of
Theorems~\ref{theorem:hcorrectness}--\ref{theorem:qcorrectness}, this will
entail that for refining maps the homographic and quadratic algorithms
correspond to \mobius and quadratic maps on \Ibase. This is the final step in
the correctness proof of the algorithm. We call this the \emph{topological
  correctness} of the algorithms. The reason is that (1) it shows that
\refiningvoid which is a purely metric property is enough to ensure the
correctness and the (2) proofs are based on continuity arguments.  It will also
show that our type theoretic approach of dealing with general corecursion has
been sound with respect to the metric semantics of the algorithms.

%%%%%%%%%%%%%%%%%%%%%%%%%%%%%%%%%%%%%%%%%%%%%%%%%%%%%%%%%%%%%%%%%%%%%%%%%%%%%%%%
%                                                                              %
%                      Homographic Topological Correctness                     %
%                                                                              %
%%%%%%%%%%%%%%%%%%%%%%%%%%%%%%%%%%%%%%%%%%%%%%%%%%%%%%%%%%%%%%%%%%%%%%%%%%%%%%%%
\subsection{Homographic Algorithm}\label{subsec:homographic-topological-correctness}
The productivity predicate $P_{h}$ is the latest in a chain of type theoretic
auxiliary predicates and functions $E_{h},\overline{m}_h$ and $\Psi_h$.  By
examining these predicates we observe that the only topological notion appears
in the type of the constructors of $E_{h}$, in the form of the emission
condition.  We should follow this link to obtain the productivity predicate for
a refining \mobius map.

First we state some elementary properties of the interval predicates that we
introduced in Section~\ref{sec:algorithms}. We omit the proofs which are trivial
case distinctions on comparisons of the end points of the intervals.
\begin{lem}\label{lemma:incl_properties}
Let $\mu$ be a \mobius map.
\begin{enumerate}[\em(1)]
\item $\mu$ is bounded (\ie its denominator does not vanish in \Ibase) if and
  only if the property \bounded{\mu} holds. \label{item:incl_properties_1}
\item  $\mu$ is refining (\ie it maps \Ibase into itself) if and only if \refining{\mu} holds. \label{item:incl_properties_2}
\item If $\mu$ is bounded and for each $r\in\Ibase$, $\mu(r)\in[-1,0]$ then \emitting{\mu}{\LL}.\label{item:incl_properties_3}
\item If $\mu$ is bounded and for each $r\in\Ibase$, $\mu(r)\in[0,1]$ then \emitting{\mu}{\RR}.\label{item:incl_properties_4}
\item If $\mu$ is bounded and for each $r\in\Ibase$, $\mu(r)\in[\frac{-1}{3},\frac{1}{3}]$ then
  \emitting{\mu}{\MM}.\label{item:incl_properties_5}\qed
\end{enumerate}
\end{lem} 

This lemma ensures us that we can comfortably work with the predicates
\refiningvoid and \boundedvoid to prove results about refining maps. An easily
provable consequence of the above lemma is the following.

\begin{lem} \label{lemma:bounded_refining_product}
Let $\mu_1$ and $\mu_2$ be \mobius maps.
\begin{enumerate}[\em(1)]
\item If \bounded{\mu_1} and \refining{\mu_2} then \bounded{\mu_1\after \mu_2}\enspace.\label{item:bounded_refining_product_1}
\item If \refining{\mu_1} and \refining{\mu_2} then \refining{\mu_1\after \mu_2}\enspace. \label{item:bounded_refining_product_2}\qed
\end{enumerate}
\end{lem}

Given a refining \mobius map $\mu$, we define the \emph{diameter} of $\mu$ to be
\begin{displaymath}
  \diameter{\mu}=|\mu(-1)-\mu(1)|\enspace.
\end{displaymath}

Next we need a metric property of the representation, which measures the
amount of the redundancy of the representation. For any set of $\Phi$ of
refining \mobius maps we define
\begin{equation*}
  \redun{\Phi} = \min\set{|\phi_i(-1)-\phi_j(1)| | \phi_i,\phi_j\in\DIG,\; \phi_i(-1)\neq\phi_j(1)}.
\end{equation*}
The above definition is based on the intuitive idea that the more overlap
between ranges of the digits, the more choices one has for representing real
numbers.  The intended meaning is that for two digit sets $\Phi_1$ and $\Phi_2$,
with the same number of elements, if $\redun{\Phi_1} > \redun{\Phi_2} $ then
$\Phi_1$ has more redundancy. Note that this intended meaning does not work for
adding extra digits (which decreases \redunvoid) but rather for comparing the
redundancy of two digits sets with the same number of digits.

Clearly
\begin{displaymath}
  \redun{\LL,\RR,\MM}=\frac{1}{3}\enspace.
\end{displaymath}

Then using this quantity we state and prove the following lemma that shows that
for refining \mobius maps with sufficiently small diameter the emission
condition holds.
\begin{lem} \label{lemma:diameter_redundancy}
  If $\mu$ is a refining \mobius map and $\diameter{\mu}<\frac{1}{3}=\redun{\LL,\RR,\MM}$ then there
  exists $\phi\in\DIG$ such that $\emitting{\mu}{\phi}$.
\end{lem}
\proof The proof uses
Lemma~\ref{lemma:incl_properties}.\ref{item:incl_properties_3}--\ref{item:incl_properties_5},
and case distinction on comparison of $\mu(-1)$ and $\mu(1)$ with
$\LL(1)=\RR(-1)=0,\MM(-1)=\frac{-1}{3}$ and $\MM(1)=\frac{1}{3}$.  \qed

At this point the following question arises: can we decrease the diameter of an
arbitrary refining \mobius maps by repeated absorption in a way that it becomes
less than ${1}/{3}$?  The answer is positive. First we should assess the
diameter of the product of two \mobius maps because an absorption step is
nothing but a product with a digit.  Hence we prove the following lemma.

\begin{lem} \label{lemma:diam_product}
Let $\mu_1=\slft a b c d$ be a refining \mobius map.
\begin{enumerate}[\em(1)]
\item  If  $\mu_2$ is a refining \mobius map, then 
\begin{displaymath}
\diameter{\mu_1\after\mu_2}= \frac{\diameter{\mu_2}\cdot|\det \mu_1|}{|(c{\mu_{2}}(-1)+d)(c{\mu_{2}}(1)+d)|}\enspace.
\end{displaymath} \label{item:diam_product_1}
\item If $\alpha\in\DIGinf$, then
\begin{displaymath}
\diameter{\mu{\after}\alpha\sb{0}{\after}\ldots{\after}\alpha\sb{n-1}}= \frac{(u_{n}(\alpha)-l_{n}(\alpha))\cdot|\det \mu_1|}{|(c\cdot u_n(\alpha)+d)(c\cdot l_n(\alpha)+d)|}\enspace.
\end{displaymath} \label{item:diam_product_2}
\end{enumerate}
\end{lem}
\proof
\begin{enumerate}[(1)]
\item Note that by
  Lemma~\ref{lemma:bounded_refining_product}.\ref{item:bounded_refining_product_2}
  the product is refining and thus the left hand side is well-defined. The
  identity follows by straightforward calculation (See~\cite{heckmann.02}).
\item Since all the digits are refining, we can apply
  part~(\ref{item:diam_product_1}) with
  $\mu_2:=\alpha\sb{0}{\after}\ldots{\after}\alpha\sb{n-1}$.\qed
\end{enumerate}

Next we can prove that after finitely many absorption steps the emitting
condition holds and thus the algorithm is `informally' productive.

\begin{thm} \label{theorem:refining_exists_n_emitting} Let $\mu=\slft a b c d$
  be a refining \mobius map and $\alpha\in\DIGinf$.  Then there exist $n\in\NN$ and
  $\phi\in\DIG$ such that
  \emitting{\mu{\after}\alpha\sb{0}{\after}\ldots{\after}\alpha\sb{n-1}}{\phi} holds.
\end{thm}
\proof Let $X:=\max(\frac{1}{|c+d|},\frac{1}{|d-c|})$ and take 
\begin{equation} \label{eq:refining_exists_n_emitting_0}
n:=\lceil {6\cdot|\det
  \mu|\cdot X^2}\rceil
\end{equation}
(here we take the ceiling using the Archimedean property of \QQ).

Note that since $\mu$ is refining then it is bounded and hence $\slft 0 1 c d$
is bounded and monotone in \Ibase.  Thus for all $x\in\Ibase$ we have
\begin{equation} \label{eq:refining_exists_n_emitting_1}
\frac{1}{|cx+d|}\leq X\enspace.
\end{equation}

On that account we calculate:
\begin{alignat*}{2}
  \diameter{\mu{\after}\alpha\sb{0}{\after}\ldots{\after}\alpha\sb{n-1}}&= \frac{(u_{n}(\alpha)-l_{n}(\alpha))\cdot|\det \mu|}{|(c\cdot u_n(\alpha)+d)(c\cdot l_n(\alpha)+d)|}  && \qquad\text{by Lemma~\ref{lemma:diam_product}.\ref{item:diam_product_2}}\\
  &\leq X^2\cdot (u_n(\alpha)-l_n(\alpha)) \cdot |\det \mu| &&\qquad \text{by~(\ref{eq:refining_exists_n_emitting_1})}\\
  &\leq \frac{2\cdot X^2\cdot|\det \mu|}{n+1}  &&\qquad \text{by~(\ref{eq:rep_3})}\\
  &< \frac{1}{3} &&\qquad \text{by~(\ref{eq:refining_exists_n_emitting_0}).}
\end{alignat*}
Hence we can apply Lemma~\ref{lemma:diameter_redundancy} and obtain $\phi$ as
required.  \qed

The above existential theorem gives us a pair of witnesses $\pair{n}{\phi}$, but
we would like to make the canonical choice of the \emph{smallest} such witness.
This is possible because we are dealing with a decidable predicate \emittingvoid
(see \emittingdecvoid), on the well-founded set $\NN\times\DIG$.  The idea is
that once we have a witness we can perform a search bounded by this witness to
obtain the smallest witness.  This can be summarised as the following result.

\begin{lem} \label{lemma:refining_exists_LNP} Let $\mu$ be a refining \mobius
  map and $\alpha\in\DIGinf$.  Then there exist $n\in\NN$ and $\phi\in\DIG$ such
  that the following two conditions hold.
  \begin{enumerate}[\em(1)]
  \item \emitting{\mu{\after}\alpha\sb{0}{\after}\ldots{\after}\alpha\sb{n-1}}{\phi}\enspace;\label{item:refining_exists_LNP_0}
  \item $\forall m<n\forall \phi', \neg
    \emitting{\mu{\after}\alpha\sb{0}{\after}\ldots{\after}\alpha\sb{m-1}}{\phi'}\enspace$. \label{item:refining_exists_LNP_1}
  \end{enumerate}
\end{lem}

At this point we are ready to embark on proving our type theoretic predicates.
First we prove that being refining implies that $E\sb{h}$ holds.

\begin{lem} \label{lemma:refining_emits} 
  Let $\mu$ be a refining \mobius map and
  $\alpha\in\DIGinf$. Then $E\sb{h}(\mu,\alpha)$ holds.
\end{lem}
\proof Assume $n$ is obtained by applying Lemma~\ref{lemma:refining_exists_LNP}
to $\mu$ and $\alpha$.  We proceed by \emph{induction}\footnote{This might seem
  odd, as $n$ is a witness given to us; nevertheless we can carry out induction
  for a universal property for any $m$ such that $m=n$.}  on $n$. If $n=0$ then
\emitting{\mu}{\phi} should hold for some $\phi$ and we can apply the
corresponding constructor of $E\sb{h}$ among $E\sb{hL},E\sb{hR}$ and $E\sb{hM}$.

Now assume we have proven the conclusion for all refining maps for which the
witness given by Lemma~\ref{lemma:refining_exists_LNP} is $k$ and $n=k+1$. Note
that for $\mu\after\alpha\sb{0}$ and $\alpha|\sb{1}$ the witness given by
Lemma~\ref{lemma:refining_exists_LNP} must be $k$. Therefore by induction
hypothesis we have
\begin{equation}\label{eq:refining_emits_1}
  E\sb{h}(\mu\after\alpha\sb{0},\alpha|\sb{1})\enspace.
\end{equation}

Since $0<k$ we know by
Lemma~\ref{lemma:refining_exists_LNP}.\ref{item:refining_exists_LNP_1} that
\emitting{\mu}{\phi} does not hold for any $\phi$, i.e.

\begin{equation}\label{eq:refining_emits_2}
  \neg\emitting{\mu}{\LL}\enspace,\quad\neg\emitting{\mu}{\RR}\enspace,\quad\neg\emitting{\mu}{\MM}\enspace.
\end{equation}

Consequently, we can apply the constructor $E\sb{hab}$ to
(\ref{eq:refining_emits_2}) and (\ref{eq:refining_emits_1}) to obtain a proof of
$E\sb{h}(\mu,\alpha)$.\qed

Next we need two technical lemmas for relating the refining property with the
two auxiliary functions $\overline{\delta}\sb{h}$
(Section~\ref{subsec:homographic-coind-correctness}) and $\overline{m}\sb{h}$
(Section~\ref{subsec:general-corec-homographic}).

\begin{lem} \label{lemma:refining_depth} 
  Let $\mu$ be a refining \mobius map and
  $\alpha\in\DIGinf$. Let $n$ be given by applying Lemma~\ref{lemma:refining_exists_LNP}
  to $\mu$ and $\alpha$. Let $t_0$ be the proof given by
  Lemma~\ref{lemma:refining_emits} that $E\sb{h}(\mu,\alpha)$ holds. Then
  \begin{displaymath}
    \overline{\delta}\sb{h}(\mu,\alpha,t_0)=n\enspace.
  \end{displaymath}
\end{lem}
\proof By induction on $n$. For $n=0$ by
Lemma~\ref{lemma:refining_exists_LNP}.\ref{item:refining_exists_LNP_0}
\emitting{\mu}{\alpha} should hold for some $\phi$, and the conclusion follows
from the definition of $\overline{\delta}\sb{h}$.

Now assume we have proven the conclusion for all refining maps for which the
witness given by Lemma~\ref{lemma:refining_exists_LNP} is $k$ and $n=k+1$.
Applying the induction hypothesis to $\mu\after\alpha\sb{0}$ and $\alpha|\sb{1}$
we obtain
\begin{equation}\label{eq:refining_depth_1}
  \overline{\delta}\sb{h}(\mu\after\alpha\sb{0},\alpha|\sb{1},t_1)=k\enspace,
\end{equation}
where $t_1$ is the proof given by Lemma~\ref{lemma:refining_emits} for
\begin{equation*}
  E\sb{h}(\mu\after\alpha\sb{0},\alpha|\sb{1})\enspace.
\end{equation*}

Furthermore since $0<k$ by
Lemma~\ref{lemma:refining_exists_LNP}.\ref{item:refining_exists_LNP_1}
\begin{equation*}
  \neg\emitting{\mu}{\LL}\enspace,\quad\neg\emitting{\mu}{\RR}\enspace,\quad\neg\emitting{\mu}{\MM}\enspace.
\end{equation*}

From here together with the definition of $\overline{\delta}\sb{h}$ and
(\ref{eq:refining_depth_1}) we obtain
$$%\begin{displaymath}
  \overline{\delta}\sb{h}(\mu,\alpha,t_0)=\overline{\delta}\sb{h}(\mu\after\alpha\sb{0},\alpha|\sb{1},t_1)+1=k+1=n\enspace.\eqno{\qEd}
$$%\end{displaymath}\qed

\begin{lem} \label{lemma:refining_modulus} Let $\mu$ be a refining \mobius map
  and $\alpha\in\DIGinf$. Let $t_0$ be the proof given by
  Lemma~\ref{lemma:refining_emits} that $E\sb{h}(\mu,\alpha)$ holds.  Let
  $\overline{m}\sb{h}(\mu,\alpha,t_0):=\pair{\phi}{\pair{\mu'}{\alpha'}}$. Then
  $\mu'$ is refining.
\end{lem}
\proof Let $n$ be obtained by applying Lemma~\ref{lemma:refining_exists_LNP} to
$\mu$ and $\alpha$.  By Lemma~\ref{lemma:refining_depth} we have
\begin{equation} \label{eq:refining_modulus_1}
  \overline{\delta}\sb{h}(\mu,\alpha,t_0)=n\enspace.
\end{equation}
Hence by applying Lemma~\ref{lemma:delta_h_modulus_h} to $\mu,\alpha,t_0$ and
$n$ we obtain a digit $\phi'$ such that
\begin{equation*} \label{eq:refining_modulus_2}
  \overline{m}\sb{h}(\mu,\alpha,t_0)=\pair{\phi'}{\pair{\phi'\sp{-1}{\after}\mu{\after}\alpha\sb{0}{\after}\ldots{\after}\alpha\sb{n-1}}{\alpha|\sb{n}}}
\end{equation*}
Hence 
\begin{equation*} \label{eq:refining_modulus_3}
  \phi'=\phi\enspace,\qquad \mu'=\phi\sp{-1}{\after}\mu{\after}\alpha\sb{0}{\after}\ldots{\after}\alpha\sb{n-1}\enspace.
\end{equation*}
Note that the inverse of digits are \emph{not} refining, so we cannot use
Lemma~\ref{lemma:bounded_refining_product} to prove that $\mu'$ is refining. But
instead we apply Lemma~\ref{lemma:refining-inverse}.  So we have to show that
\emitting{{\mu\sp{\prime}}}{\phi}. But this is evident by applying
Lemma~\ref{lemma:delta_h_emttingvoid} to~(\ref{eq:refining_modulus_1}).  \qed

Finally, we prove that being refining implies $\Psi\sb{h}$, which is the crux of
the productivity property that we used for defining the homographic algorithm.

\begin{lem} \label{lemma:refining_Psi}
  Let $n\in\NN$, $\mu$ be a refining \mobius map and $\alpha\in\DIGinf$.
  Then $\Psi\sb{h}(n,\mu,\alpha)$ holds.
\end{lem}
\proof By induction on $n$. If $n=0$ then by Lemma~\ref{lemma:refining_emits} we
know that $E\sb{h}(\mu,\alpha)$ holds, and hence we can apply the constructor
$\Psi\sb{h0}$ to obtain the conclusion. Assume the conclusion holds for $n=k$
and arbitrary refining \mobius maps. Let $t_0$ be the \emph{specific} proof
given by Lemma~\ref{lemma:refining_emits} that $E\sb{h}(\mu,\alpha)$ holds, and
let
\begin{displaymath}
  \overline{m}\sb{h}(\mu,\alpha,t_0):=\pair{\phi}{\pair{\mu'}{\alpha'}}\enspace.
\end{displaymath}
Due to our choice of $t_0$, by Lemma~\ref{lemma:refining_modulus} it follows
that $\mu'$ is refining.  Thus we can apply the induction hypothesis to $\mu'$
and $\alpha'$ to obtain a proof of $\Psi\sb{h}(k,\mu',\alpha')$ which can be
rewritten as:
\begin{equation} \label{eq:refining_Psi_1}
  \Psi\sb{h}\big(k,\pi\sb{23}(\overline{m}\sb{h}(\mu,\alpha,t_0)),\pi\sb{33}(\overline{m}\sb{h}(\mu,\alpha,t_0))\big)\enspace.
\end{equation}
Hence by applying the constructor $\Psi\sb{hS}$ to $\mu,\alpha, t_0$ and
(\ref{eq:refining_Psi_1}) the result follows.  \qed

As a corollary we obtain the main result of this section. This states that the
purely topological property \refiningvoid entails the type theoretic
productivity predicate $P\sb{h}$ which we had added to satisfy the guardedness
condition of \coq.
\begin{cor} \label{corollary:refining_productivity}
  Let $\mu$ be a refining \mobius map and $\alpha\in\DIGinf$. Then
  $P\sb{h}(\mu,\alpha)$ holds.
\end{cor}

%%%%%%%%%%%%%%%%%%%%%%%%%%%%%%%%%%%%%%%%%%%%%%%%%%%%%%%%%%%%%%%%%%%%%%%%%%%%%%%%
%                                                                              %
%                       Quadratic Topological Correctness                      %
%                                                                              %
%%%%%%%%%%%%%%%%%%%%%%%%%%%%%%%%%%%%%%%%%%%%%%%%%%%%%%%%%%%%%%%%%%%%%%%%%%%%%%%%
\subsection{Quadratic Algorithm}\label{subsec:quadratic-topological-correctness}
We should prove that if a quadratic map is refining then the predicates
$E\sb{q},\Psi\sb{q}$ and $P\sb{q}$ hold. We follow the same route as for the
homographic algorithm to prove the counterpart of
Theorem~\ref{theorem:refining_exists_n_emitting}. There we defined the diameter
of a refining \mobius map and calculated a uniform upper bound for it after
finite absorption steps. Here the situation is slightly more complicated,
because when assessing the effect of a quadratic map on two intervals (one for
each argument) we should examine the values at 4 corners of the Cartesian
product. So already the definition of the diameter will be slightly different.
But first we state the properties of the interval predicates for a quadratic map
(see Appendix~\ref{appendix:quadratic-predicates} for the definitions), which
are again provable by straightforward case analysis.

\begin{lem}\label{lemma:incl_properties_quadratic}
Let $\xi$ be a quadratic map.
\begin{enumerate}[\em(1)]
\item $\xi$ is bounded (\ie its denominator does not vanish in
  $\Ibase\times\Ibase$) if and only if \bounded{\xi}
  holds.\label{item:incl_properties_quadratic_1}
\item $\xi$ is refining (\ie it maps $\Ibase\times\Ibase$ into itself) if and
  only if \refining{\xi} holds.\label{item:incl_properties_quadratic_2}
\item If $\xi$ is bounded and for each $r_1,r_2\in\Ibase$, $\xi(r1,r2)\in[-1,0]$
  then \emitting{\xi}{\LL}.\label{item:incl_properties_quadratic_3}
\item If $\xi$ is bounded and for each $r_1,r_2\in\Ibase$, $\xi(r1,r2)\in[0,1]$
  then \emitting{\xi}{\RR}.\label{item:incl_properties_quadratic_4}
\item If $\xi$ is bounded and for each $r_1,r_2\in\Ibase$,
  $\xi(r1,r2)\in[\frac{-1}{3},\frac{1}{3}]$ then
  \emitting{\xi}{\MM}.\label{item:incl_properties_quadratic_5}\qed
\end{enumerate}
\end{lem} 

Recall that there were two ways of composing a quadratic map and a \mobius map.
Using the lemma above we can derive the following about these two products.

\begin{lem} \label{lemma:bounded_refining_product_quadratic} Let $\xi$ be a
  quadratic map and $\mu_1$ and $\mu_2$ be \mobius maps.
\begin{enumerate}[\em(1)]
\item If \bounded{\xi}, \refining{\mu_1} and \refining{\mu_2} then \bounded{\xi
    \afterL \mu_1\afterR \mu_2}\enspace. \label{item:bounded_refining_product_quadratic_1}
\item If \refining{\xi}, \refining{\mu_1} and \refining{\mu_2} then
  \refining{\xi \afterL \mu_1\afterR \mu_2}\enspace.
  \label{item:bounded_refining_product_quadratic_2}\qed
\end{enumerate}
\end{lem}

Note that for a refining quadratic map $\xi$, $\xi(\Ibase,\_)$ is a function on
subintervals of \Ibase.  We define the \emph{diameter} of $\xi$ on
\emph{rational} subintervals $[x_0,y_0]$ and $[x_1,y_1]$ of \Ibase to be
\begin{displaymath}
\tdiameter{\xi,[x_0,y_0],[x_1,y_1]}=[\min(r_{00},r_{01},r_{10},r_{11}),\max(r_{00},r_{01},r_{10},r_{11})]\enspace,
\end{displaymath}
where 
\begin{displaymath}
\slft {r_{00}} {r_{01}} {r_{10}} {r_{11}} = \slft {\xi(x_0,y_0)} {\xi(x_0,y_1)} {\xi(x_1,y_0)} {\xi(x_1,y_1)}\enspace.
\end{displaymath}

We will usually use \tdiameter{\xi,\Ibase,\Ibase} however in some intermediate
steps of the proofs we sometimes have to invoke diameter for other rational
subintervals. Again we can prove a lemma relating the diameter, redundancy and
\emittingvoid.

\begin{lem} \label{lemma:diameter_redundancy_quadratic} If $\xi$ is a refining
  quadratic map for which
  $\tdiameter{\xi,\Ibase,\Ibase}<\frac{1}{3}=\redun{\LL,\RR,\MM}$ then there
  exists $\phi\in\DIG$ such that $\emitting{\xi}{\phi}$.
\end{lem}
\proof
If $\xi(\Ibase,\Ibase)=[x,y]$, we consider the following three cases.
\begin{align*}
  \text{If } y\leq 0:&\qquad\qquad \emitting{\xi}{\LL}&\quad\text{by Lemma~\ref{lemma:incl_properties_quadratic}.\ref{item:incl_properties_quadratic_3}}\enspace,\\
  \text{else if } 0\leq x:&\qquad\qquad \emitting{\xi}{\RR}&\quad\text{by Lemma~\ref{lemma:incl_properties_quadratic}.\ref{item:incl_properties_quadratic_4}}\enspace,\\
  \text{otherwise}:&\qquad\qquad \emitting{\xi}{\MM}&\quad\text{by
    Lemma~\ref{lemma:incl_properties_quadratic}.\ref{item:incl_properties_quadratic_5}}\enspace.\rlap{\hbox to 55 pt{\hfill\qEd}}
\end{align*}

Unlike what we did in Lemma~\ref{lemma:diam_product}, here we cannot find a
closed formula for the diameter of the product of a quadratic map and two
\mobius maps. We should find another way of ensuring that in the absorption
steps the diameter can become smaller than $1/3$. At the first glance it seems
that we really have to prove the uniform continuity of quadratic map considered
as a binary function on rational numbers, but careful examination of the proof
of Theorem~\ref{theorem:refining_exists_n_emitting} shows that the pointwise
continuity could be enough. Thus we prove the following lemma.

\begin{lem} \label{lemma:quadratic_map_continuous} Let $\xi$ be a refining
  quadratic map. Then for all $0<\varepsilon\in\QQ^{+}$ there exist
  $0<\vartheta_0,\vartheta_1\in\QQ^{+}$ such that for all
  $x_0,x_1,y_0,y_1\in\Ibase$ if $|x_0-x_1|<\vartheta_0$ and
  $|y_0-y_1|<\vartheta_1$ then
  \begin{displaymath}
    |\xi(x_0,y_0)-\xi(x_1,y_1)|<\varepsilon\enspace.
  \end{displaymath}
\end{lem}
\proof This is equivalent to the continuity of a refining (and
hence bounded) quadratic map on $\Ibase\times\Ibase$. 
\qed

As a corollary we can locally bound the diameter of a refining quadratic map:

\begin{cor} \label{corollary:diameter2_bounded_quadratic} Let $\xi$ be a
  refining quadratic map. Then for all $0<\varepsilon\in\QQ^{+}$ there exist
  $0<\vartheta_0,\vartheta_1\in\QQ^{+}$ such that for all $x_0,x_1,y_0,y_1\in\Ibase$
  if $|x_0-x_1|<\vartheta_0$ and $|y_0-y_1|<\vartheta_1$ then
  \begin{displaymath}
    \tdiameter{\xi,[x_0,y_0],[x_1,y_1]}<\varepsilon\enspace.
  \end{displaymath}
\end{cor}

At this point we are ready to state and prove the counterpart of
Theorem~\ref{theorem:refining_exists_n_emitting}, that ensures the flow of
emission steps after a finite number of absorption steps.

\begin{thm} \label{theorem:refining_exists_n_emitting_quadratic} Let $\xi$ be a
  refining quadratic map and $\alpha,\beta\in\DIGinf$.  Then there exist $n\in\NN$
  and $\phi\in\DIG$ such that
  \emitting{\xi\pair{\alpha\sb{0}{\after}\ldots{\after}\alpha\sb{n-1}}{\beta\sb{0}{\after}\ldots{\after}\beta\sb{n-1}}}{\phi}
  holds.
\end{thm}
\proof Let $\vartheta_0,\vartheta_1$ be given by applying
Corollary~\ref{corollary:diameter2_bounded_quadratic} to $\varepsilon=1/3$. Take
\begin{displaymath}
  n:=\max(\lceil \frac{2}{\vartheta_0}\rceil,\lceil\frac{2}{\vartheta_1}\rceil)\enspace.
\end{displaymath}

Let 
\begin{displaymath}
  x_0:=l_n(\alpha)\enspace,\quad y_0:=l_n(\beta)\enspace,\quad x_1:=u_n(\alpha)\enspace,\quad y_1:=u_n(\beta)\enspace,
\end{displaymath}

Note that due to~(\ref{eq:rep_3}) we have 
\begin{equation*} \label{eq:refining_exists_n_emitting_quadratic_1}
|x_0-x_1|\leq\frac{2}{n+1}\enspace,\qquad |y_0-y_1|\leq\frac{2}{n+1}
\end{equation*}

From here together with Corollary~\ref{corollary:diameter2_bounded_quadratic}
for $x_0,x_1,y_0$ and $y_1$ we obtain
\begin{equation}\label{eq:refining_exists_n_emitting_quadratic_2}
  \tdiameter{\xi,[x_0,y_0],[x_1,y_1]}<\frac{1}{3}\enspace.
\end{equation}

But an easy calculation shows that 
\begin{align}\label{eq:refining_exists_n_emitting_quadratic_3}
  \tdiameter{\xi,[x_0,y_0],[x_1,y_1]}&=\tdiameter{\xi,\alpha\sb{0}{\after}\ldots{\after}\alpha\sb{n-1}\Ibase,\beta\sb{0}{\after}\ldots{\after}\beta\sb{n-1}\Ibase}\notag\\
  &=\tdiameter{\xi\pair{\alpha\sb{0}{\after}\ldots{\after}\alpha\sb{n-1}}{\beta\sb{0}{\after}\ldots{\after}\beta\sb{n-1}},\Ibase,\Ibase}\enspace.
\end{align}
Therefore we can apply Lemma~\ref{lemma:diameter_redundancy_quadratic} with
(\ref{eq:refining_exists_n_emitting_quadratic_2}) and
(\ref{eq:refining_exists_n_emitting_quadratic_3}) to obtain the desired digit
$\phi$.  \qed

The above proof is based on the pointwise continuity of refining quadratic maps.
Of course we can find a uniform bound to be applied in the proof of the
algorithm, but that would require a formalisation of the bivariate version of
the Heine--Borel theorem.  For our purpose the above proof suffices, because it
gives us a witness which we will use in a bounded search for finding the
\emph{smallest} such witness.  Furthermore, the pointwise continuity gives a
finer estimate of the complexity of the algorithm, but we will not pursue this
matter here.  Results concerning the complexity of these algorithms can be found
in~\cite{heckmann.98,krznaric.00}.

The discrepancy between the homographic algorithm and the quadratic algorithm
ends here. This means that the remaining steps for deriving the productivity
predicate $P\sb{q}$ for the refining quadratic maps is essentially the same as
those for the homographic algorithm. Therefore we only present the three
important statements here, and we refrain from repeating the arguments. The
proofs can be consulted in the formalisation package~\cite{milad.07}.

\begin{lem} \label{lemma:refining_emits_quadratic}
  Let $\xi$ be a refining quadratic map and $\alpha,\beta\in\DIGinf$. Then
  $E\sb{q}(\xi,\alpha,\beta)$ holds.
\end{lem}

\begin{lem} \label{lemma:refining_Psi_quadratic} Let $n\in\NN$, $\xi$ be
  a refining quadratic map and $\alpha,\beta\in\DIGinf$.  Then
  $\Psi\sb{q}(n,\xi,\alpha,\beta)$ holds.
\end{lem}

\begin{cor} \label{corollary:refining_productivity_quadratic} Let $\xi$ be a
  refining quadratic map and $\alpha,\beta\in\DIGinf$. Then
  $P\sb{q}(\xi,\alpha,\beta)$ holds.
\end{cor}

As expected the productivity predicate $P\sb{q}$, being a \emph{local} property
of the domain of the algorithm gives a finer description of the domain of the
algorithm.  For example the addition tensor is not refining on \Ibase but the
quadratic algorithm applied with the addition tensor is productive for
$\alpha,\beta\in[-\frac{1}{4},\frac{1}{4}]$. However, if we transfer the
computations to the entire real line by adding a redundant sign bit, then the
refining quadratic maps are enough for calculating elementary
functions~\cite{potts.98}.

%%%%%%%%%%%%%%%%%%%%%%%%%%%%%%%%%%%%%%%%%%%%%%%%%%%%%%%%%%%%%%%%%%%%%%%%%%%%%%%%
%                                                                              %
%                           Constructive or Classical                          %
%                                                                              %
%%%%%%%%%%%%%%%%%%%%%%%%%%%%%%%%%%%%%%%%%%%%%%%%%%%%%%%%%%%%%%%%%%%%%%%%%%%%%%%%
\section{Reexamining The Method}
We can outline the path that we followed in this article in the following steps.
\begin{enumerate}[(1)]
\item Implementing algorithms in type theoretic language.  \label{item:method_1}
\item Proving that they satisfy their \haskell-like
  specification.\label{item:method_2}
\item Proving that the algorithms correspond to \emph{partial} \mobius and
  quadratic maps on \Ibase:\label{item:method_3}
\begin{enumerate}[(a)]
\item They are total on those subsets of $\MAT{\times}\DIGinf$ and
  $\TEN{\times}\DIGinf$ for which the productivity predicates $P\sb{h}$ and
  $P\sb{q}$ hold.\label{item:method_4}
\item They are total on the subsets of $\MAT$ and $\TEN$ containing refining
  maps.\label{item:method_5}
\end{enumerate}
\end{enumerate}

There are different aspects of the proofs in each of these phases that we would
like to clarify.

\subsection*{Dependence on Representation}
None of the steps above \emph{depend} on the representation, although the
specific proofs about this representation, as well as our choice of the base
interval appeared frequently in our reasoning. In \cite[Chapter~5]{milad.04b} we
show that as long as a representation satisfies a few properties with respect to
the effect of its digits on the chosen base interval, it will not affect the
productivity and thus correctness behaviour of the algorithms.  For example for
any such representation and for any choice of base interval we can derive a
counterpart of (\ref{eq:rep_3}) --- with a different bound--- and
Lemma~\ref{lemma:diam_product}.\ref{item:diam_product_1}.

\subsection*{Type Theoretic vs. Topological Properties}
We pointed out this correlation throughout the article. Here we summarise it for
all the above steps. Step~\ref{item:method_1} above is purely type theoretic. It
simply consist of writing a function parametrised by a proof obligation that
passes the type checking in the functional programming language of \coq. The
proofs (proof irrelevance lemmas and termination certificates) are objects that
are meaningful and expressible in a framework where dependent types and
(co)inductive types exist. These proofs ensure an initial layer of correctness:
that the input and output have the right type and whether the algorithms are
productive for given inputs. One could say that in this step the domain and
codomain of the algorithms are described.

Step~\ref{item:method_2} is the first phase in proving correctness with respect
to the intended semantics, but it still has a purely type theoretic nature. The
proofs are based on working with bisimulation and deriving cofixed point
equations for the algorithms. They relate the algorithms to their specification
written in a more liberal functional programming (\eg Haskell), where
termination and productivity are not hampering us.

Step~\ref{item:method_3} possesses both type-theoretic and topological elements.
In Step~\ref{item:method_4} this is best captured in the relationship between
\rep (a type theoretic predicate) and \decode{\_} (a purely topological
operation).  While in Step~\ref{item:method_5} the topological aspects dealing
with the pointwise continuity of the underlying maps are dominant.  Still we can
observe that
Lemmas~\ref{lemma:refining_emits}--\ref{corollary:refining_productivity} resort
to type theoretic properties such as proof irrelevance and termination
certificates. The correlation of these two aspects is highlighted in the final
propositions about the purely metric property \refiningvoid and the productivity
predicates that were required for the type-checking in Step~\ref{item:method_1}.

\subsection*{Statistics on Formalisation}
Finally we present some of the statistics pertaining to the formalised
algorithms.  They indicate the size (in kilobytes) and length (in number of
lines\footnote{Number of lines is obtained using the command \texttt{coqwc}
  which disregards the commented and blank lines.}) of the ASCII code of the
formalisation.

In Table~\ref{table:statistics_1} we present the relative size of the
formalisation work for each of the above steps. We also add a separate category
(last row) for parts of the formalisation that included general results about
digits, \mobius maps, quadratic maps and several interval predicates.  In
Table~\ref{table:statistics_2} we take an alternative viewpoint and present the
statistics for each algorithm separately. The first row (digits) denotes the
part that was common to both algorithms. The last column gives the size of the
\haskell program that is obtained by using the \emph{program extraction
  mechanism} of \coq~\cite[\S 18]{coq}.

\begin{table}[h] 
  \begin{tabular}{|l|l|l|}
    \hline
    Step& Size (percentage of total) & Length  \\
    \hline
    Step~\ref{item:method_1} & 27 K (5.5\%) & 529 lines  \\
    Step~\ref{item:method_2} & 8 K (1.6\%) & 166 lines  \\
    Step~\ref{item:method_4}& 111 K (23.4\%) & 2099 lines  \\
    Step~\ref{item:method_5}& 62 K (12.9\%) & 1107 lines  \\
    General facts& 268 K (56.3\%)& 4596 lines  \\
    \emph{Total}  &  475 K (100\%)&  8497 lines \\
    \hline
  \end{tabular}
  \refstepcounter{tableau}
  \label{table:statistics_1} 
  \caption{\textbf{Various phases of the formalisation.}}
\end{table}

\begin{table}[h] 
  \begin{tabular}{|l|l|l|l|}
    \hline
    task& size & lines & {extracted \haskell} \\
    \hline
    {digits}& 128 K & 2541 lines & 51 lines \\
    {homographic}& 147 K & 2845 lines & 61 lines \\
    quadratic& 200 K & 3111 lines & 126 lines \\
    \emph{Total}  &  475 K &  8497 lines &  238 lines\\
    \hline
  \end{tabular}
  \refstepcounter{tableau}
  \label{table:statistics_2} 
  \caption{\textbf{Relative size of proofs and programs for different algorithms.}}
\end{table}

The presented statistics provide a good indication of the state of the art in
formalising mathematical results and verifying algorithms. However, (and
fortunately) these statistics are likely to be outdated as new versions of \coq
featuring more automation tools will become available.

%%%%%%%%%%%%%%%%%%%%%%%%%%%%%%%%%%%%%%%%%%%%%%%%%%%%%%%%%%%%%%%%%%%%%%%%%%%%%%%%
%                                                                              %
%                           Section: Conclusion                                %
%                                                                              %
%%%%%%%%%%%%%%%%%%%%%%%%%%%%%%%%%%%%%%%%%%%%%%%%%%%%%%%%%%%%%%%%%%%%%%%%%%%%%%%%
\section{Conclusions and Further Work}\label{sec:conclusion}
We have shown the correctness of the homographic and quadratic algorithms on a
stream representation of real numbers in \Ibase. Following the general set-up of
\cite[\S~5]{milad.04b} the method is easily extensible to any admissible digit
set for any compact proper subinterval of the extended real numbers
$[-\infty,+\infty]$. Our correctness proofs use an inductive productivity
predicate and a coinductive predicate \rep that relates \DIGinf and \Ibase.  We
use the coinductive machinery of the \coq proof assistant to formalise functions
on infinite objects and coinductive proofs. In particular we base our treatment
of coinductive functions on their cofixed point equations. These exploit the
inherent infinite nature of streams by adhering to \bisim which is a
bisimulation relation and is more suitable than the inductive (\emph{Leibniz})
equality. The coinductive arguments themselves are independent of \coq and can
be formalised in any proof assistant that accommodates coinductive types.
Furthermore we prove --- in \coq--- that for the class of \mobius and quadratic
maps that satisfy a metric property (being refining), the homographic and
quadratic algorithms will output provably infinite streams for any input.

Among several perceivable directions for the future work, the more immediate one
would be to continue the \coq formalisation of the algorithms, by developing a
fully modular framework that axiomatises the properties of representations and
refining maps that are needed for the formalisation. Each specific
representation would then be portable into our formalisation if a suitable
interface is satisfied. This will pave the way for applying our formalisation to
more efficient representations such as the one used by
Edalat--Potts~\cite{edalat.97} or the ternary one used in
\cite{ciaffaglione.06,hou.06,bertot.06,marcial.07}.

The big picture would be to continue working on the formalisation of the
Edalat--Potts framework for exact real arithmetic.  The homographic and
quadratic algorithm are the base case of Edalat and Potts' \emph{normalisation
  algorithm} which is defined on the coinductive type of expression
trees~\cite{edalat.97,potts.98}.  Therefore if we could apply the method of this
article to formalise and verify this algorithm we could obtain all the
elementary functions. Unfortunately this does not seem to be possible: the
difficulty lies in the general corecursion used in the normalisation algorithm,
The method of the Section~\ref{sec:general-corec} needs a more complicated
machinery than that of \CIC to be applicable to the normalisation algorithm.
This is because the normalisation algorithm is a nested algorithm and therefore
applying our method the modulus of productivity $\overline{m}\sb{h}$ will be a
nested function too.  It is well-known that applying Bove--Capretta method for
formalising nested recursive functions requires the presence of
\emph{inductive--recursive} types~\cite{bove.01,dybjer.00}. In this case the
inductive domain predicate will become an inductive--recursive predicate that is
defined simultaneously with the nested function. A similar phenomenon happens in
our method, in the sense that we need a notion similar to induction--recursion
that would allow for the simultaneous definition of an inductive predicate
together with a cofixed point.  The author is exploring the possibility of
defining such a notion.  Recent work by Setzer on combining induction--recursion
and general recursion seems to open up new possibilities for our work in this
directions~\cite{setzer.06}. 

One can contemplate of adapting and generalising our method for lazy exact
arithmetic algorithms beyond the Edalat--Potts algorithm. One starting point in
this direction would be to follow the technique used in ~\cite{simpson.98} (see
also Section \ref{subsec:general-corec?})  for obtaining higher order functions
on real numbers such as the numerical integration.

\section*{Acknowledgement}
The author wishes to thank the anonymous referees for their valuable comments
and their proposed simplifications that helped improve the paper.

\newcommand{\noopsort}[1]{}

\newpage
\appendix
\section{Interval Predicates for Quadratic Algorithm}\label{appendix:quadratic-predicates}
Let $\xi=\stensor a b c d e f g h$ be a quadratic map and $\phi=\slft
{\phi_{00}} {\phi_{01}} {\phi_{00}} {\phi_{01}} \in \DIG$. Then:

\begin{align*}
  \bounded{\xi} &:= (0{<}e{+}f{+}g{+}h \wedge 0{<}e{-}f{-}g{+}h \wedge 0{<}{-}e{-}f{+}g{+}h \wedge 0{<}{-}e{+}f{-}g{+}h \bigvee\\
  &\quad\qquad e{+}f{+}g{+}h{<}0\wedge e{-}f{-}g{+}h{<}0\wedge{-}e{-}f{+}g{+}h{<}0\wedge{-}e{+}f{-}g{+}h{<}0)\enspace;\\
\end{align*}

\begin{align*}
  \refining{\xi} &:= \bounded{\xi} \bigwedge\\
  &\quad\quad \Big{(}0{\leq} a{+}b{+}c{+}d{+}e{+}f{+}g{+}h \wedge 0{\leq}{-}a{-}b{-}c{-}d{+}e{+}f{+}g{+}h \wedge\qquad\qquad\qquad\qquad\\
  &\quad\qquad 0{\leq} a{-}b{-}c{+}d{+}e{-}f{-}g{+}h \wedge 0{\leq}{-}a{+}b{+}c{-}d{+}e{-}f{-}g{+}h \wedge\\
  &\quad\qquad 0{\leq}{-}a{-}b{+}c{+}d{-}e{-}f{+}g{+}h \wedge 0{\leq} a{+}b{-}c{-}d{-}e{-}f{+}g{+}h \wedge\\
  &\quad\qquad 0{\leq}{-}a{+}b{-}c{+}d{-}e{+}f{-}g{+}h \wedge 0{\leq} a{-}b{+}c{-}d{-}e{+}f{-}g{+}h \bigvee\\
  &\quad\qquad a{+}b{+}c{+}d{+}e{+}f{+}g{+}h {\leq}0\wedge {-}a{-}b{-}c{-}d{+}e{+}f{+}g{+}h {\leq}0\wedge\\
  &\quad\qquad a{-}b{-}c{+}d{+}e{-}f{-}g{+}h {\leq}0\wedge {-}a{+}b{+}c{-}d{+}e{-}f{-}g{+}h {\leq}0\wedge\\
  &\quad\qquad {-}a{-}b{+}c{+}d{-}e{-}f{+}g{+}h {\leq}0\wedge a{+}b{-}c{-}d{-}e{-}f{+}g{+}h {\leq}0\wedge\\
  &\quad\qquad {-}a{+}b{-}c{+}d{-}e{+}f{-}g{+}h {\leq}0\wedge a{-}b{+}c{-}d{-}e{+}f{-}g{+}h {\leq}0\Big{)}\enspace;\\
\end{align*}

\begin{align*}
\emitting{\xi}{\phi} &:= \bounded{\xi} \wedge \\
&(e{-}f{-}g{+}h)(e{-}f{-}g{+}h)({\phi_{01}}{-}{\phi_{00}}){\leq}
(e{-}f{-}g{+}h)(a{-}b{-}c{+}d)({\phi_{11}}{-}{\phi_{10}}) \wedge\\
&(e{-}f{-}g{+}h)(a{-}b{-}c{+}d)({\phi_{10}}{+}{\phi_{11}}){\leq}
(e{-}f{-}g{+}h)(e{-}f{-}g{+}h)({\phi_{00}}{+}{\phi_{01}}) \wedge\\
&({-}e{-}f{+}g{+}h)({-}e{-}f{+}g{+}h)({\phi_{01}}{-}{\phi_{00}}){\leq}
({-}e{-}f{+}g{+}h)({-}a{-}b{+}c{+}d)({\phi_{11}}{-}{\phi_{10}}) \wedge\\
&({-}e{-}f{+}g{+}h)({-}a{-}b{+}c{+}d)({\phi_{10}}{+}{\phi_{11}}){\leq}
({-}e{-}f{+}g{+}h)({-}e{-}f{+}g{+}h)({\phi_{00}}{+}{\phi_{01}}) \wedge\\
&({-}e{+}f{-}g{+}h)({-}e{+}f{-}g{+}h)({\phi_{01}}{-}{\phi_{00}}){\leq}
({-}e{+}f{-}g{+}h)({-}a{+}b{-}c{+}d)({\phi_{11}}{-}{\phi_{10}}) \wedge\\
&({-}e{+}f{-}g{+}h)({-}a{+}b{-}c{+}d)({\phi_{10}}{+}{\phi_{11}}){\leq}
({-}e{+}f{-}g{+}h)({-}e{+}f{-}g{+}h)({\phi_{00}}{+}{\phi_{01}}) \wedge\\
&(e{+}f{+}g{+}h)(e{+}f{+}g{+}h)({\phi_{01}}{-}{\phi_{00}}){\leq}
(e{+}f{+}g{+}h)(a{+}b{+}c{+}d)({\phi_{11}}{-}{\phi_{10}}) \wedge\\
&(e{+}f{+}g{+}h)(a{+}b{+}c{+}d)({\phi_{10}}{+}{\phi_{11}}){\leq}
(e{+}f{+}g{+}h)(e{+}f{+}g{+}h)({\phi_{00}}{+}{\phi_{01}})\enspace.
\end{align*}

\section{Correspondence with the formalised \coq files.} \label{appendix:names_formalised} 

In the following table we present the correspondence between the terms and
lemmas in the article and their formalised version in~\cite{milad.07}. In the
second column \texttt{foo.bar} refers to the \coq term \texttt{bar} in file
\texttt{foo.v} which is available for public download at~\cite{milad.07}. Note
that for notations that are overloaded between the homographic and quadratic
case, in the first column we explicitly mention an argument. For brevity we drop
this argument to the \coq functions; \eg \bounded{\mu} is in fact formalised as
\texttt{digits.Bounded\_M~mu} but we ignore \texttt{mu}.
\begin{longtable}[c]{|l|l|}
  \hline
  \textbf{Item in article}& \textbf{Formalised Version}\\ 
  \hline
  \endhead
  \hline
  \endfoot
  \bisim & \texttt{digits.bisim}   \\
  \bounded{\mu} & \texttt{digits.Bounded\_M}\\
  \bounded{\xi} & \texttt{digits.Bounded\_T}\\
  \refining{\mu} & \texttt{digits.Is\_refining\_M}\\
  \refining{\xi} & \texttt{digits.Is\_refining\_T}\\
  \DIG & \texttt{digits.Digit}\\
  \LL  & \texttt{digits.LL}\\
  \RR  & \texttt{digits.RR}\\
  \MM  & \texttt{digits.MM}\\
  \DIGinf & \texttt{digits.Reals}\\
  \emitting{\mu}{\_} & \texttt{digits.Incl\_M}\\
  \emitting{\xi}{\_} & \texttt{digits.Incl\_T}\\
  \ensuremath{\after} (two matrices) & \texttt{digits.product}\\
  \afterM (matrix and tensor) & \texttt{digits.m\_product}\\
  \afterL & \texttt{digits.left\_product}\\
  \afterR & \texttt{digits.right\_product}\\
  \(E\sb{h}\) & \texttt{homographic.emits\_h}\\
  \(\overline{m}\sb{h}\) & \texttt{homographic.modulus\_h}\\
  \emittingdec{\mu}{\_} & \texttt{digits.Incl\_M\_dec\_D}\\
  \(E\sb{hab}\_\textrm{inv}\) & \texttt{homographic.emits\_h\_absorbs\_inv}\\
  Lemma~\ref{lemma:modulus_h_PI}%3.1%
                                    & \texttt{homographic.modulus\_h\_PI} \\
  Lemma~\ref{lemma:modulus_h_fixedpoints}.\ref{item:modulus_h_fixedpoints_1}%3.2.1%
                                    & \texttt{homographic.modulus\_h\_L} \\
  Lemma~\ref{lemma:modulus_h_fixedpoints}.\ref{item:modulus_h_fixedpoints_2}%3.2.2%
                                    & \texttt{homographic.modulus\_h\_R} \\
  Lemma~\ref{lemma:modulus_h_fixedpoints}.\ref{item:modulus_h_fixedpoints_3}%3.2.3%
                                    & \texttt{homographic.modulus\_h\_M} \\
  Lemma~\ref{lemma:modulus_h_fixedpoints}.\ref{item:modulus_h_fixedpoints_4}%3.2.4%
                                    & \texttt{homographic.modulus\_h\_absorbs} \\
  \(\Psi\sb{h}\) & \texttt{homographic.step\_productive\_h}\\
  \(P\sb{h}\)& \texttt{homographic.productive\_h}\\
  \(P\sb{h}\_E\sb{h}\) & \texttt{homographic.productive\_h\_emits\_h}\\
  \(\overline{m}\sb{h}\_P\sb{h}\) & \texttt{homographic.modulus\_h\_productive\_h}\\
  Lemma~\ref{lemma:Psi_h_inv}.\ref{item:Psi_h_inv_1}%3.3.1%
                                    & \texttt{homographic.step\_productive\_h\_inv\_1} \\
  Lemma~\ref{lemma:Psi_h_inv}.\ref{item:Psi_h_inv_2}%3.3.2%
                                    & \texttt{homographic.step\_productive\_h\_inv\_2} \\
  \(\bar{h}\) & \texttt{homographic.homographic}\\
  Lemma~\ref{lemma:homographic_EPI}%3.4%
                                    & \texttt{homographic.homographic\_EPI} \\
  Lemma~\ref{lemma:homographic_cofixedpoints}.\ref{item:homographic_cofixedpoints_1}%3.5.1%
                                    & \texttt{homographic.homographic\_emits\_L} \\
  Lemma~\ref{lemma:homographic_cofixedpoints}.\ref{item:homographic_cofixedpoints_2}%3.5.2%
                                    & \texttt{homographic.homographic\_emits\_R} \\
  Lemma~\ref{lemma:homographic_cofixedpoints}.\ref{item:homographic_cofixedpoints_3}%3.5.3%
                                    & \texttt{homographic.homographic\_emits\_M} \\
  Lemma~\ref{lemma:homographic_cofixedpoints}.\ref{item:homographic_cofixedpoints_4}%3.5.4%
                                    & \texttt{homographic.homographic\_absorbs} \\
  \(E\sb{q}\) & \texttt{quadratic.emits\_q}\\
  \(\overline{m}\sb{q}\) & \texttt{quadratic.modulus\_q}\\
  \emittingdec{\xi}{\_} & \texttt{digits.Incl\_T\_dec\_D}\\
  \(E\sb{qab}\_\textrm{inv}\) & \texttt{quadratic.emits\_q\_absorbs\_inv}\\
  \(\Psi\sb{q}\) & \texttt{quadratic.step\_productive\_q}\\
  \(P\sb{q}\)& \texttt{quadratic.productive\_q}\\
  \(P\sb{q}\_E\sb{q}\) & \texttt{quadratic.productive\_q\_emits\_q}\\
  \(\overline{m}\sb{q}\_P\sb{q}\) & \texttt{quadratic.modulus\_q\_productive\_q}\\
  \(\bar{q}\) & \texttt{quadratic.quadratic}\\
  Lemma~\ref{lemma:quadratic_EPI}%3.6%
                                    & \texttt{quadratic.quadratic\_EPI} \\
  Lemma~\ref{lemma:quadratic_cofixedpoints}.\ref{item:quadratic_cofixedpoints_1}%3.7.1%
                                    & \texttt{quadratic.quadratic\_emits\_L} \\
  Lemma~\ref{lemma:quadratic_cofixedpoints}.\ref{item:quadratic_cofixedpoints_2}%3.7.2%
                                    & \texttt{quadratic.quadratic\_emits\_R} \\
  Lemma~\ref{lemma:quadratic_cofixedpoints}.\ref{item:quadratic_cofixedpoints_3}%3.7.3%
                                    & \texttt{quadratic.quadratic\_emits\_M} \\
  Lemma~\ref{lemma:quadratic_cofixedpoints}.\ref{item:quadratic_cofixedpoints_4}%3.7.4%
                                    & \texttt{quadratic.quadratic\_absorbs} \\
  \rep & \texttt{rep.rep}\\
  (\ref{eq:rep_1}) & \texttt{rep.rep\_stepl} \\
  (\ref{eq:rep_10}) & \texttt{rep.rep\_L\_inv} \\
  $\alpha|_{n}$ & \texttt{Streams\_addenda.drop}\\
  $\alpha_{n-1}^{-1}\after\ldots \alpha_{0}^{-1}$ & \texttt{digits.product\_init\_rev}\\
  (\ref{eq:rep_2}) & \texttt{rep.rep\_drop} \\
  \decode{\_} & \texttt{Cauchy\_stream.real\_value}\\
  (\ref{eq:rep_3}) & \texttt{ub.thesis\_5\_7\_9} \\
  (\ref{eq:rep_11}) (lower bound) & \texttt{Cauchy\_stream.rep\_lb} \\
  (\ref{eq:rep_11}) (upper bound)& \texttt{Cauchy\_stream.rep\_ub} \\
  (\ref{eq:rep_8}) & \texttt{rep.rep\_inv\_interval} \\
  (\ref{eq:rep_4}) & \texttt{Cauchy\_stream.real\_value\_digits} \\
  (\ref{eq:rep_5}) & \texttt{Cauchy\_stream.real\_value\_base\_interval} \\
  (\ref{eq:rep_9}) & \texttt{Cauchy\_stream.rep\_real\_value} \\
  (\ref{eq:rep_7}) & \texttt{Cauchy\_stream.real\_value\_implies\_rep}\\
  (\ref{eq:rep_6}) & \texttt{Cauchy\_stream.rep\_implies\_real\_value}\\
  \(\overline{\delta}\sb{h}\) & \texttt{hcorrectness.depth\_h}\\
  \ensuremath{\mu{\after}\alpha\sb{0}{\after}\ldots{\after}\alpha\sb{n-1}} & \texttt{hcorrectness.product\_init}\\
  Lemma~\ref{lemma:delta_h_modulus_h} & \texttt{hcorrectness.depth\_h\_modulus\_h}\\
  Lemma~\ref{lemma:delta_h_emttingvoid} & \texttt{hcorrectness.depth\_h\_Incl\_M\_inf\_strong}\\
  Lemma~\ref{lemma:homographic_emitting} & \texttt{hcorrectness.homographic\_emits\_strong}\\
  Lemma~\ref{lemma:refining-inverse} & \texttt{Refining\_M.Incl\_M\_absorbs\_Is\_refining\_M}\\
  Theorem~\ref{theorem:hcorrectness} & \texttt{hcorrectness.homographic\_correctness}\\
  \(\overline{\delta}\sb{q}\) & \texttt{qcorrectness.depth\_q}\\
  \ensuremath{\xi\pair{\alpha\sb{0}{\after}\ldots{\after}\alpha\sb{n-1}}{\beta\sb{0}{\after}\ldots{\after}\beta\sb{n-1}}} & \texttt{qcorrectness.product\_init\_zip}\\
  Lemma~\ref{lemma:quadratic_emitting} & \texttt{qcorrectness.quadratic\_emits\_strong}\\
  Lemma~\ref{lemma:quadratic-refining-inverse} & \texttt{Refining\_T.Incl\_T\_absorbs\_Is\_refining\_T}\\
  Theorem~\ref{theorem:qcorrectness} & \texttt{qcorrectness.quadratic\_correctness}\\
  Lemma~\ref{lemma:incl_properties}.\ref{item:incl_properties_1} ($\Rightarrow$) & \texttt{Bounded\_M.denom\_nonvanishing\_M\_Bounded\_M}\\
  Lemma~\ref{lemma:incl_properties}.\ref{item:incl_properties_1} ($\Leftarrow$) & \texttt{Bounded\_M.Bounded\_M\_denom\_nonvanishing\_M}\\
  Lemma~\ref{lemma:incl_properties}.\ref{item:incl_properties_2} ($\Rightarrow$) & \texttt{Refining\_M.Is\_refining\_M\_property\_fold}\\
  Lemma~\ref{lemma:incl_properties}.\ref{item:incl_properties_2} ($\Leftarrow$) & \texttt{Refining\_M.Is\_refining\_M\_property}\\
  Lemma~\ref{lemma:incl_properties}.\ref{item:incl_properties_3} & \texttt{Incl\_M.Incl\_M\_L\_folded}\\
  Lemma~\ref{lemma:incl_properties}.\ref{item:incl_properties_4} & \texttt{Incl\_M.Incl\_M\_R\_folded}\\
  Lemma~\ref{lemma:incl_properties}.\ref{item:incl_properties_5} & \texttt{Incl\_M.Incl\_M\_M\_folded}\\
  Lemma~\ref{lemma:bounded_refining_product}.\ref{item:bounded_refining_product_1} & \texttt{Refining\_M.Is\_refining\_M\_Bounded\_M\_product}\\
  Lemma~\ref{lemma:bounded_refining_product}.\ref{item:bounded_refining_product_2} & \texttt{Refining\_M.Is\_refining\_M\_product}\\
  \diameter{\mu} & \texttt{digits.diameter} \\
  \redunvoid & \texttt{digits.redundancy} \\
  Lemma~\ref{lemma:diameter_redundancy} & \texttt{productivity\_M.thesis\_5\_6\_9}\\
  \ensuremath{\slft{0}{1}{c}{d}} (for \ensuremath{\slft{a}{b}{c}{d}}) & \texttt{digits.eta\_discriminant}\\
  Lemma~\ref{lemma:diam_product}.\ref{item:diam_product_1} & \texttt{productivity\_M.diameter\_product}\\
  Lemma~\ref{lemma:diam_product}.\ref{item:diam_product_2} & \texttt{productivity\_M.diameter\_product\_init}\\
  Theorem~\ref{theorem:refining_exists_n_emitting} & \texttt{productivity\_M.thesis\_5\_6\_10}\\
  Lemma~\ref{lemma:refining_exists_LNP} & \texttt{productivity\_M.semantic\_modulus\_h}\\
  Lemma~\ref{lemma:refining_emits} & \texttt{productivity\_M.Is\_refining\_M\_emits\_h}\\
  Lemma~\ref{lemma:refining_depth} & \texttt{productivity\_M.Is\_refining\_M\_depth\_h}\\
  Lemma~\ref{lemma:refining_modulus} & \texttt{productivity\_M.Is\_refining\_M\_modulus\_h} \\
  Lemma~\ref{lemma:refining_Psi} & \texttt{productivity\_M.Is\_refining\_M\_step\_productive\_h}\\
  Corollary~\ref{corollary:refining_productivity} & \texttt{productivity\_M.Is\_refining\_M\_productive\_h}\\
  Lemma~\ref{lemma:incl_properties_quadratic}.\ref{item:incl_properties_quadratic_1} ($\Rightarrow$) & \texttt{Bounded\_T.denom\_nonvanishing\_T\_Bounded\_T} \\
  Lemma~\ref{lemma:incl_properties_quadratic}.\ref{item:incl_properties_quadratic_1} ($\Leftarrow$) & \texttt{Bounded\_T.Bounded\_T\_denom\_nonvanishing\_T} \\
  Lemma~\ref{lemma:incl_properties_quadratic}.\ref{item:incl_properties_quadratic_2} ($\Rightarrow$) & \texttt{Refining\_T.Is\_refining\_T\_property\_fold} \\
  Lemma~\ref{lemma:incl_properties_quadratic}.\ref{item:incl_properties_quadratic_2} ($\Leftarrow$) & \texttt{Refining\_T.Is\_refining\_T\_property} \\
  Lemma~\ref{lemma:incl_properties_quadratic}.\ref{item:incl_properties_quadratic_3} & \texttt{Incl\_T.Incl\_T\_L\_folded} \\
  Lemma~\ref{lemma:incl_properties_quadratic}.\ref{item:incl_properties_quadratic_4} & \texttt{Incl\_T.Incl\_T\_R\_folded} \\
  Lemma~\ref{lemma:incl_properties_quadratic}.\ref{item:incl_properties_quadratic_5} & \texttt{Incl\_T.Incl\_T\_M\_folded} \\
  Lemma~\ref{lemma:bounded_refining_product_quadratic}.\ref{item:bounded_refining_product_quadratic_1} & \texttt{Refining\_T.}\\ 
 & \hfill\texttt{Is\_refining\_T\_Bounded\_T\_left\_right\_product} \\
  Lemma~\ref{lemma:bounded_refining_product_quadratic}.\ref{item:bounded_refining_product_quadratic_2} & \texttt{Refining\_T.Is\_refining\_T\_left\_right\_product} \\
  \tdiameter{\xi} & \texttt{digits.diameter2} \\
  Lemma~\ref{lemma:diameter_redundancy_quadratic} & \texttt{productivity\_T.thesis\_5\_6\_20} \\
  Lemma~\ref{lemma:quadratic_map_continuous} & \texttt{productivity\_T.upper\_bound\_diameter2} \\
  Corollary~\ref{corollary:diameter2_bounded_quadratic} & \texttt{productivity\_T.thesis\_5\_6\_19} \\
  Theorem~\ref{theorem:refining_exists_n_emitting_quadratic} & \texttt{productivity\_T.thesis\_5\_6\_10'} \\
  Lemma~\ref{lemma:refining_emits_quadratic} & \texttt{productivity\_T.Is\_refining\_T\_emits\_q} \\
  Lemma~\ref{lemma:refining_Psi_quadratic} &  \texttt{productivity\_T.Is\_refining\_T\_step\_productive\_q} \\
  Corollary~\ref{corollary:refining_productivity_quadratic} & \texttt{productivity\_T.Is\_refining\_T\_productive\_q} \\
 \hline
\end{longtable}

\end{document}